\documentclass[reprint,superscriptaddress,amsmath, amssymb,aps,prd,notitlepage,longbibliography,floatfix,nofootinbib,onecolumn]{revtex4-1}

\usepackage{amsmath}
\usepackage{lipsum}
\allowdisplaybreaks[4]
\usepackage{cancel}
\usepackage{extarrows}
\usepackage{tensor}     
\usepackage{float}
\usepackage[final]{graphicx}   
\usepackage[
colorlinks=true,        
citecolor=blue,         
linkcolor=blue,         
urlcolor=blue           
]{hyperref}             
\usepackage{bm}         
\usepackage{xcolor}     
\usepackage{lipsum}
\usepackage{color}      
\usepackage[utf8]{inputenc} 
\usepackage[section]{placeins} 
\usepackage{appendix}
\usepackage{units}
\usepackage[capitalise]{cleveref}
\usepackage{booktabs}
\usepackage{multirow}
\usepackage{import}
\usepackage{adjustbox}
\usepackage{url}
\newcommand{\nc}{\newcommand*}

\nc{\xbar}{\bar{x}}
\nc{\rhoeq}{\rho_{\mathrm{eq}}}
\nc{\zeq}{z_{\mathrm{eq}}}
\nc{\tla}{\tilde{\lambda}}
\nc{\bt}{\beta}
\nc{\dt}{\delta}
\nc{\Dt}{\Delta}
\nc{\vj}{\vec{j}}
\nc{\vl}{\vec{l}}
\nc{\hx}{\hat{x}}
\nc{\hy}{\hat{y}}
\nc{\bj}{\bm{j}}
\nc{\mJ}{\mathcal{J}}
\nc{\mP}{\mathcal{P}}
\nc{\Msun}{M_\odot}
\nc{\app}{\approx}
\nc{\av}[1]{\langle #1 \rangle}
\nc{\eq}[1]{Eq.~\eqref{#1}}
\nc{\al}{\alpha}
\nc{\Xstar}{X_{\ast}}
\nc{\fpbh}{f_{\mathrm{pbh}}}
\nc{\vth}{\vec{\theta}}
\nc{\vla}{\vec{\lambda}}
\nc{\vd}{\vec{d}}
\nc{\Mmin}{M_{\mathrm{min}}}
\nc{\rmd}{\mathrm{d}}
\nc{\mmin}{{m_{\mathrm{min}}}}
\nc{\mmax}{{m_{\mathrm{max}}}}
\nc{\mR}{\mathcal{R}}
\nc{\tmR}{\tilde{\mathcal{R}}}
\nc{\s}{\sigma}
\nc{\ogw}{\Omega_{\mathrm{GW}}}
\nc{\addref}{[\textcolor{red}{add ref}] }
\nc{\Om}{\Omega}
\nc{\gm}{\gamma}
\nc{\Gm}{\Gamma}
\nc{\gpcyr}{\mathrm{Gpc}^{-3}\,\mathrm{yr}^{-1}}
\nc{\Eq}[1]{Eq.~\eqref{#1}}
\nc{\Fig}[1]{Fig.~\ref{#1}}
\nc{\Table}[1]{Table~\ref{#1}}
\nc{\lvc}{LIGO/Virgo} 
\nc{\Sec}[1]{Sec.~\ref{#1}}
\nc{\eg}{\textit{e.g.~}}
\nc{\SNR}{\mathrm{SNR}}
\nc{\be}{\mathbf{\epsilon}}
\nc{\bn}{\mathbf{n}}
\nc{\bd}{\mathbf{d}}
\nc{\ba}{\mathbf{a}}
\nc{\eps}{\epsilon}
\nc{\bnu}{\mathbf{\nu}}
\nc{\mb}{\mathbf}
\nc{\bbt}{\mathbf{t}}
\nc{\bth}{\mathbf{\theta}}
\nc{\bep}{\mathbf{\epsilon}}
\nc{\uni}{\mathrm{U}}
\nc{\logu}{\operatorname{\mathrm{log-U}}}
\nc{\RN}{\mathrm{RN}}
\nc{\BN}{\mathrm{BN}}
\nc{\GN}{\mathrm{GN}}
\nc{\mcN}{\mathcal{N}}
\nc{\GWB}{\mathrm{GW}}
\nc{\yr}{\mathrm{yr}}
\nc{\Am}{\mathcal{A}}
\nc{\Dm}{\mathcal{D}}
\nc{\Hm}{\mathcal{H}}
\nc{\sovast}{Soviet Ast.}

\nc{\kmsmpc}{\mathrm{km\ s^{-1} Mpc^{-1}}}
\nc{\lcdm}{\Lambda\mathrm{CDM}}
\nc{\ev}{\mathrm{eV}}

\nc{\mrm}{\mathrm}
\nc{\BE}{B\scriptsize{AYES}\normalsize{E}\scriptsize{PHEM}\normalsize  }

\nc{\Ostgw}{\Omega_{\mathrm{GW}}^{\mathrm{ST}}}
\nc{\Ottgw}{\Omega_{\mathrm{GW}}^{\mathrm{TT}}}
\nc{\Ovlgw}{\Omega_{\mathrm{GW}}^{\mathrm{VL}}}
\nc{\Oslgw}{\Omega_{\mathrm{GW}}^{\mathrm{SL}}}
\nc{\cosxi}{\beta}

\nc{\gmPL}{\gamma_{\mathrm{PL}}}
\nc{\APL}{A_{\mathrm{PL}}}

\def\({\left(}
\def\){\right)}
\def\[{\left[}
\def\]{\right]}

\def\e{\begin{equation}}
\def\q{\end{equation}}
\def\m{\begin{eqnarray}}
\def\n{\end{eqnarray}}
\nc{\red}[1]{\textcolor{red}{#1}}

\begin{document}

\title{Cosmological Constraints on Neutrino Mass within Consistent Cosmological Models}

\author{Ye-Huang Pang}
\email{pangyehuang22@mails.ucas.ac.cn}
\affiliation{School of Fundamental Physics and Mathematical Sciences, Hangzhou Institute for Advanced Study, UCAS, Hangzhou 310024, China}
\affiliation{School of Physical Sciences, 
    University of Chinese Academy of Sciences, 
    No. 19A Yuquan Road, Beijing 100049, China}
\affiliation{CAS Key Laboratory of Theoretical Physics, 
    Institute of Theoretical Physics, Chinese Academy of Sciences,Beijing 100190, China}
\author{Xue Zhang}
\email{zhangxue@yzu.edu.cn}
\affiliation{Center for Gravitation and Cosmology, 
    College of Physical Science and Technology, 
    Yangzhou University, Yangzhou 225009, China}
\author{Qing-Guo Huang}
\email{huangqg@itp.ac.cn}
\affiliation{School of Fundamental Physics and Mathematical Sciences, Hangzhou Institute for Advanced Study, UCAS, Hangzhou 310024, China}
\affiliation{School of Physical Sciences, 
    University of Chinese Academy of Sciences, 
    No. 19A Yuquan Road, Beijing 100049, China}
\affiliation{CAS Key Laboratory of Theoretical Physics, 
    Institute of Theoretical Physics, Chinese Academy of Sciences,Beijing 100190, China}


\begin{abstract}


Recently, the emergence of cosmological tension has raised doubts about the consistency of the $\lcdm$ model. 
In order to constrain the neutrino mass within a consistent cosmological framework, we investigate three massive neutrinos with normal hierarchy (NH) and inverted hierarchy (IH) in both the axion-like EDE (Axi-EDE) model and the AdS-EDE model.
We use the joint datasets including cosmic microwave background (CMB) power spectrum from \textit{Planck} 2018, Pantheon of type Ia supernova, baryon acoustic oscillation (BAO) and $H_0$ data from SH0ES. 
For the $\nu$Axi-EDE model, we obtain $\sum m_{\nu,\mathrm{NH}} < 0.152~\ev$ and $\sum m_{\nu,\mathrm{IH}} < 0.178~\ev$, while for the $\nu$AdS-EDE model, we find $\sum m_{\nu,\mathrm{NH}} < 0.135~\ev$ and $\sum m_{\nu,\mathrm{IH}} < 0.167~\ev$.
Our results exhibit a preference for the normal hierarchy in both the $\nu$Axi-EDE model and the $\nu$AdS-EDE model.

\end{abstract}
\maketitle

\section{Introduction} 
In the standard model of particle physics, three generations of neutrinos, namely $\nu_e, \nu_\mu$ and $\nu_\tau$, are associated with three neutrino flavor eigenstates. The phenomenon of neutrino oscillation provides evidence that the neutrinos are massive \cite{Super-Kamiokande:1998kpq,SNO:2002tuh}.
The neutrinos mass constraints can be derived from various particle physics experiments; however, the precise value of the neutrino mass remains unclear to date. Among these measurements, the kinematics of tritium decay can provide an upper limit on the neutrino mass \cite{KATRIN:2019yun}. Neutrino oscillation provides an approach to probe the splittings of the neutrino mass-squared $\Delta m_{ij}$. The resulting mass-squared splittings are \cite{ParticleDataGroup:2014cgo}
\begin{align}
    \Delta m_{21}^2 & \equiv m_2^2 - m_1^2 \simeq 7.5 \times 10^{-5} \mathrm{eV}^2, \\
    |\Delta m_{31}^2 | & \equiv |m_3^2 - m_1^2 | \simeq 2.5 \times 10^{-3} \mathrm{eV}^2,
\end{align}
where $m_1, m_2$ and $m_3$ are neutrino mass eigenvalues that can be transformed to flavor eigenstates with neutrino mixing matrix \cite{ParticleDataGroup:2014cgo}.
The sign of $\Delta m_{31}^2$ remains undetermined, thus the above results present two possibilities of neutrino mass hierarchy: the normal hierarchy (NH) characterized by $m_1<m_2<m_3$ and the inverted hierarchy (IH) characterized by $m_3<m_1<m_2$. We denote the sum of the neutrino mass as $\sum m_\nu$. In the NH case, there exists a lower limit of the total neutrino mass, which is $\sum m_{\nu,\mathrm{NH}} > 0.06 ~\ev$. Similarly, for the IH case, we have $\sum m_{\nu,\mathrm{IH}} > 0.10~\ev$.

Cosmological observations can also provide constraints on neutrino mass, particularly the total neutrino mass $\sum m_\nu$ \cite{Planck:2018vyg,Palanque-Delabrouille:2019iyz,eBOSS:2020yzd,DES:2021wwk}. 
In the early universe, massive neutrinos are ultra-relativistic, which can be regarded as radiation. However, in late-time universe, massive neutrinos become non-relativistic and behave like cold dark matter. 
The special characteristic of massive neutrino would have significant implications on cosmological evolution at both background and perturbation levels.
Consequently, this would lead to a reshaping of the CMB power spectrum \cite{Wong:2011ip,Lesgourgues:2012uu,Lesgourgues:2013sjj_Neutrino_Cosmology,Lesgourgues:2014zoa,TopicalConvenersKNAbazajianJECarlstromATLee:2013bxd}. 
To remain consistent with CMB observations, there exists an upper limit on the sum of neutrino masses $\sum m_\nu$.
It should be noted that the inference of total neutrino mass constraints from CMB is model-dependent. In concordance $\Lambda\mathrm{CDM}$ model, it is found that $\sum m_\nu<0.26$ eV at 95\% confidence level (C.L.) using Planck 2018 TT, TE, EE+lowE and assuming degeneracy hierarchy ($m_1 = m_2 = m_3$), where lowE refers to the low-$\ell$ ($2\leqslant \ell \leqslant 29$) EE polarization power spectra \cite{Planck:2018vyg}. 
Adding baryon acoustic oscillation (BAO) data results in a more stringent constraint $\sum m_\nu <0.12$ eV at 95\% C.L. 
The neutrino mass hierarchy can be distinguished through strict cosmological constraints derived from cosmological observations. The cosmological constraints on the neutrino mass hierarchy was firstly proposed in \cite{Huang:2015wrx}. Subsequently, an updated constraint was derived using Planck 2018 TT, TE, EE+lowE+BAO, yielding $\sum m_{\nu,\mathrm{NH}} <0.15$ eV and  $\sum m_{\nu,\mathrm{IH}} <0.17$ eV at 95\% C.L. \cite{RoyChoudhury:2019hls}.

The constraints on neutrino mass in alternative cosmological models beyond the standard $\Lambda$CDM model have been extensively investigated \cite{Reid:2009nq, Kumar:2016zpg, Yang:2017amu, Vagnozzi:2018jhn, Guo:2018gyo, RoyChoudhury:2018gay, Feng:2019mym, Feng:2019jqa, Li:2020gtk, Yang:2020ope, Cardona:2020ama, Gomez-Valent:2022bku, Yadav:2023qfj,Lu:2016hsd,Sekiguchi:2020igz}. 
The new physics in these models may impose distinct constraints on the neutrino mass compared to those derived from $\Lambda$CDM model. 
In recent years, a multitude of cosmological models have been proposed to alleviate the Hubble tension \cite{DiValentino:2021izs,Abdalla:2022yfr}, which refers to the discrepancy between the local distance ladder measurement and the early-time CMB measurement. 
According to the CMB observations by \textit{Planck}  \cite{Planck:2018nkj}, $H_0 = 67.27 \pm 0.60$ $\kmsmpc$ in $\lcdm$ model. 
However, the latest $H_0$ measurement by SH0ES yields $H_0 = 73.04 \pm 1.04$ $\kmsmpc$ \cite{Riess:2021jrx}.
The discrepancy between these two results reaches a $5\sigma$ significance level. Therefore, the $\lcdm$ model may potentially exhibit inconsistencies, thus we need to constrain the neutrino masses within a consistent cosmological framework. 
Early dark energy (EDE) is a promising solution to alleviate Hubble tension by increasing the Hubble rate $H(z)$ prior to recombination. It reduces the sound horizon at recombination while increasing the $H_0$ value without shifting the angular scale of the sound horizon. 
Since CMB has tight constraints on the angular scale of sound horizon, introducing EDE can increase the value of $H_0$ without violating the fit to CMB power spectrum. In recent years, several EDE models with different mechanisms have been proposed \cite{Poulin:2023lkg, Poulin:2018cxd, Lin:2019qug, Smith:2019ihp, Agrawal:2019lmo, Braglia:2020bym, Niedermann:2019olb, Ye:2020btb, Niedermann:2020dwg, Karwal:2021vpk, McDonough:2021pdg}, 
such as the axion-like EDE model (Axi-EDE) \cite{Poulin:2018cxd, Smith:2019ihp} and the EDE model with an anti-de Sitter phase around recombination (AdS-EDE) \cite{Ye:2020btb}.
In this work, we aim to constrain neutrino mass within Axi-EDE model and AdS-EDE model,
while investigating the preferred hierarchy based on current observational data.

This paper is organized as follows. In \Sec{data}, we introduce the datasets and methods for constraining neutrino mass. Models and their parameters are summarized in \Sec{models}. The results of parameter constraints and discussions are presented in \Sec{results}. Finally, our conclusions are outlined in \Sec{conclusions}.

\section{Datasets}\label{data}

We use the combination of following datasets to constrain the neutrino mass and other cosmological parameters.

\begin{itemize}
    \item CMB
    
    CMB data include \textit{Planck} 2018 high-$\ell$ \texttt{Plik} likelihood of TT power spectrum ($30\leqslant\ell \lesssim 2500$), TE cross correlation power spectrum and EE power spectrum ($30\leqslant\ell \lesssim 2000$) \cite{Planck:2019nip}; low-$\ell$ TT and EE power spectrum ($2\leqslant\ell \leqslant 30$) and lensing power spectra \cite{Planck:2018lbu}.
    
    \item SN Ia
    
    We use the data of 1048 type Ia supernova in redshift range $0.01<z<2.3$ from Pantheon \cite{Pan-STARRS1:2017jku} which provides the information of distance module versus redshift for each supernova.
    
    \item BAO

    The BAO data include the $r_d/D_V$ value of 6dFGS at $z_\mathrm{eff} = 0.15$ \cite{Beutler:2011hx}, 
    the $D_V/r_d$ value of SDSS DR7 at $z_\mathrm{eff} = 0.15$ \cite{Ross:2014qpa}  and $D_A/r_d$, $Hr_d$ at $z_{\mathrm{eff}} = 0.38, 0.51, 0.61$ from
    BOSS DR12 \cite{BOSS:2016wmc}, where $r_d$ is the sound horizon at baryon drag epoch, 
    \begin{equation}
        r_d = \int_{z_d}^{\infty} \dfrac{c_s(z)}{H(z)} \mathrm{d}z,
    \end{equation}
    and $c_s(z)$ is the sound speed of the baryon-photon fluid. 
    The angular diameter distance $D_A$  is given by
    \begin{equation}
        D_A(z) = \dfrac{1}{1+z} \int_{0}^{z} \dfrac{\mathrm{d} z^\prime}{H(z^\prime)}.
    \end{equation}
    Then the definition of $D_V$ is
    \begin{equation}
        D_V(z) = \left[(1+z)^2 D_A^2(z) \dfrac{z}{H(z)}\right]^{1/3}.
    \end{equation}
    
    \item $H_0$ measurement
    
    Supernovae and $H_0$ for the Equation of State of dark energy (SH0ES) measures $H_0$ value through three-step local distance ladder, and the latest $H_0$ measurement result is $H_0 = 73.04 \pm 1.04 \kmsmpc$ \cite{Riess:2021jrx}. We adopt Gaussian prior of $H_0$ in our analysis.
\end{itemize}

We perform Markov-chain Monte Carlo (MCMC) sampling using the Python package \texttt{cobaya} \cite{Torrado:2020dgo}, 
and the chains is considered to be converged if the Gelman-Rubin criterion \cite{Gelman:1992zz} $R-1 < 0.05$ is satisfied.

\section{Models}\label{models}
It is argued that the activation of an axion-like scalar field $\phi$ around recombination can serve as early dark energy (EDE), potentially resolving the  $H_0$ tension \cite{Poulin:2018cxd}.
This axion-like early dark energy (denoted by Axi-EDE) can be described by a potential function
\begin{equation}
    V(\phi) = m^2 f^2 [1-\cos(\phi/f)]^n,
\end{equation}
where $n$ is a phenomenological parameter. When $n=1$, $m$ represents the axion mass and $f$ denotes the axion decay constant. In the context of alleviating $H_0$ tension, $n$ is usually chosen to be $3$. 
The field is frozen at an initial value due to Hubble friction, and rolls down the potential when the Hubble parameter decreases to a critical level. The field acts as an early dark energy component before recombination, leading to a reduction the sound horizon at recombination $r_s(z_*)$. When the field oscillates around the minimum, its equation of state change to $w_n = (n-1)/(n+1)$. Its energy density dilutes rapidly and hardly affects the subsequent cosmological evolution. 
The fractional energy density of Axi-EDE is defined as $f_{\mathrm{EDE}}(z) \equiv \rho_{\mathrm{EDE}}(z) / \rho_{\mathrm{tot}}(z)$, and the redshift at which $f_{\mathrm{EDE}}(z)$ reaches its maximum is denoted as $z_c$. For simplicity, we use $f_{\mathrm{EDE}}$ to represent $f_{\mathrm{EDE}}(z_c)$. If $m$ and $f$ are given, $f_{\mathrm{EDE}}$ and $z_c$ can be solved numerically. Thus, the Axi-EDE model has 3 additional parameters, namely $f_{\mathrm{EDE}}, z_c$ and the initial field value $\phi_i$. The robustness of the Axi-EDE model has been evaluated using various datasets \cite{Hill:2020osr,Murgia:2020ryi,Ivanov:2020ril,Poulin:2023lkg}. 
%
%
%

In \cite{Ye:2020btb}, the author proposes an early dark energy model with an AdS phase around recombination to effectively drive the values of $f_{\mathrm{EDE}}$ and $H_0$ to higher ones. In this model, the potential of the scalar field $\phi$ is 
\begin{equation}
    V(\phi) = V_0 (\phi/M_{\mathrm{Pl}})^{2n} - V_{\mathrm{AdS}},
\end{equation}
where $V_0$ is a constant, $V_{\mathrm{AdS}}$ is the depth of AdS well, and $M_{\mathrm{Pl}} = 1/\sqrt{8\pi G_N}$ is the reduced Planck mass. The potential function behaves similarly to the Rock ’n’ Roll Early Dark Energy model (RnR-EDE) \cite{Agrawal:2019lmo} when the field is allowed to drop into an AdS phase and climb out the potential well in the subsequent evolution. The AdS-EDE field is also frozen initially, and $z_c$ denotes the redshift at which the field begins rolling down.
The AdS-EDE model also introduces three extra parameters, namely $V_0, \phi_i$ and $V_{\mathrm{AdS}}$. 
These parameters are usually transformed to $f_{\mathrm{EDE}}$, $\log_{10} (1+z_c)$ and $\alpha_{\mathrm{AdS}}$. 
The relationship between $\alpha_{\mathrm{AdS}}$ and $V_{\mathrm{AdS}}$ is $V_{\mathrm{AdS}} \equiv \alpha_{\mathrm{AdS}}[\rho_m(z_c) + \rho_r(z_c)]$, where $\rho_m$ and $\rho_r$ are density of matter and radiation.
The AdS-EDE model requires a suitable selection of $f_{\mathrm{EDE}}$ and $\log_{10} (1+z_c)$ parameters, i.e. $V_{\mathrm{AdS}}$ and $V_0$), to ensure the field would not escape from the AdS potential well and prevent a collapsing universe which is undesirable. 
This feature will result in a non-zero $f_{\mathrm{EDE}}$ when constraining AdS-EDE with CMB alone \cite{Jiang:2021bab}. 
It is shown that the $H_0$ value inferred from AdS-EDE,  with fixing $\alpha_{\mathrm{AdS}} = 3.79 \times 10^{-4}$, significantly alleviates the tension \cite{Ye:2020btb,Jiang:2021bab,Ye:2021nej,Ye:2022afu}.
Note that the constraints on $\alpha_{\mathrm{AdS}}$ may be different depending on the datasets used. Therefore, we consider $\alpha_{\mathrm{AdS}}$ as a free parameter in our analysis.


It is commonly assumed that one neutrino has a mass of $m_\nu = 0.06\ev$ and the other two neutrinos are massless in cosmological models \cite{Planck:2018vyg}.
We follow this assumption for both the $\nu$Axi-EDE and $\nu$AdS-EDE base models. However, in order to constrain neutrino mass in the NH or IH case, it is necessary to consider three massive neutrino cases in $\nu$Axi-EDE and $\nu$AdS-EDE models.
%
To investigate the inclination towards NH or IH, we further use the neutrino mass hierarchy parameter defined as
\begin{equation}
    \Delta \equiv \dfrac{m_3 - m_1}{m_1 + m_3}.
\end{equation}
With this parameter, we can integrate the NH and IH into a unified model, in which $\Delta >0$ corresponds to NH and $\Delta< 0$ represents IH. Consequently, the ratio of probabilities between $\Delta >0$ and $\Delta <0$ can reflect the preference inclination.
The neutrino masses can be expressed in terms of $\Delta$ \cite{Xu:2016ddc}
\begin{align}
    m_1 & = \dfrac{1-\Delta}{2\sqrt{|\Delta|}} \sqrt{|\Delta m_{31}^2|}, \\ 
    m_2 & = \sqrt{m_1^2 + \Delta m_{21}^2}, \\
    m_3 & = \dfrac{1+\Delta}{2\sqrt{|\Delta|}} \sqrt{|\Delta m_{31}^2|}.
\end{align}

In addition to the 6 parameters in the $\lcdm$ model, the extra sampling parameters are $\{f_{\mathrm{EDE}}, \log_{10}a_c, \phi_i, m_{\min} \}$ for the NH or IH case in the $\nu$Axi-EDE model, while $\sum m_\nu$ is a derived parameter. 
For the $\nu$Axi-EDE model with $\Delta$ parameter, the extra sampling parameters are $\{f_{\mathrm{EDE}}, \log_{10}a_c, \phi_i, \Delta \}$, while $m_{\min}$ and $\sum m_\nu$ are derived parameters.
For the NH or IH case in the $\nu$AdS-EDE model, the extra sampling parameters are $\{f_{\mathrm{EDE}}, \ln (1+z_c), \alpha_{\mathrm{AdS}}, m_{\min} \}$, while $\sum m_\nu$ is derived parameter.
For $\nu$AdS-EDE model with the parameter $\Delta$, the extra sampling parameters are $\{f_{\mathrm{EDE}}, \log_{10}a_c, \alpha_{\mathrm{AdS}}, \Delta \}$, while $m_{\min}$ and $\sum m_\nu$ are derived parameters.
%
%
To calculate the theoretical prediction of observable quantities, we use the modified versions of \texttt{class} package for the $\nu$Axi-EDE model \cite{Murgia:2020ryi}\footnote{\url{https://github.com/PoulinV/AxiCLASS}} and for the $\nu$AdS-EDE model \footnote{\url{https://github.com/genye00/class_multiscf}}.

\section{Results and Discussion} \label{results}

\begin{figure}[!htbp]
    \centering
        \includegraphics[width=0.45\textwidth]{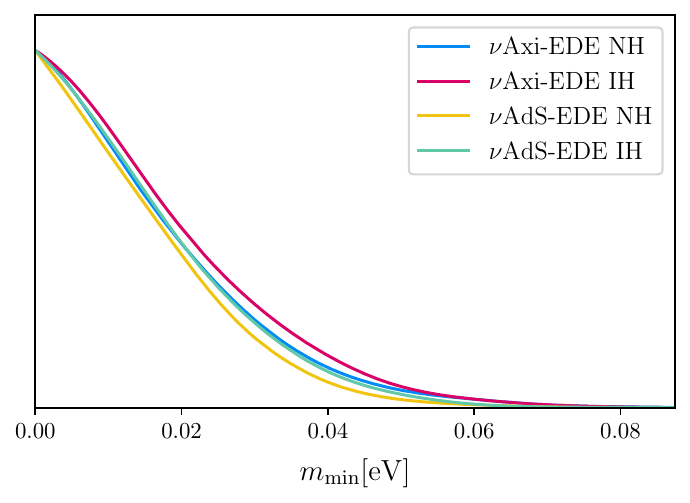}
        \includegraphics[width=0.45\textwidth]{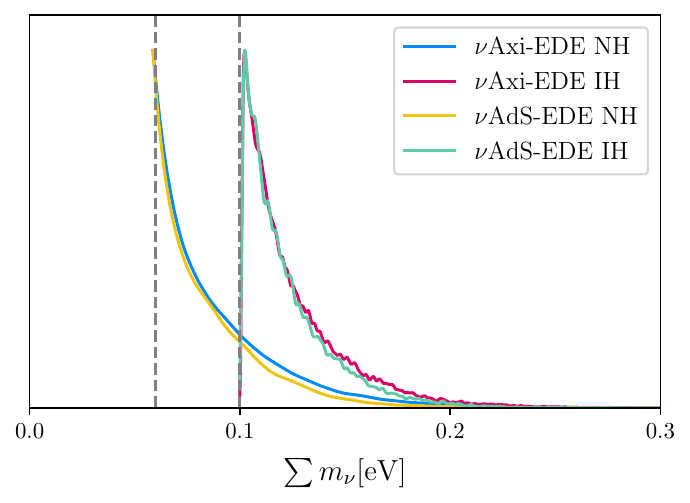}
    \caption{The 1D posterior distribution of $m_{\min}$ and $\sum m_\nu$ for the NH and IH case in the $\nu$Axi-EDE and $\nu$AdS-EDE models. The gray dashed lines in the right panel represent the lower limits of $\sum m_\nu$ in the NH case ($0.06\ev$) and the IH case ($0.1\ev$).}
    \label{fig:m}
\end{figure}

\begin{figure}[!htbp]
    \centering
    \includegraphics[width=0.5\linewidth]{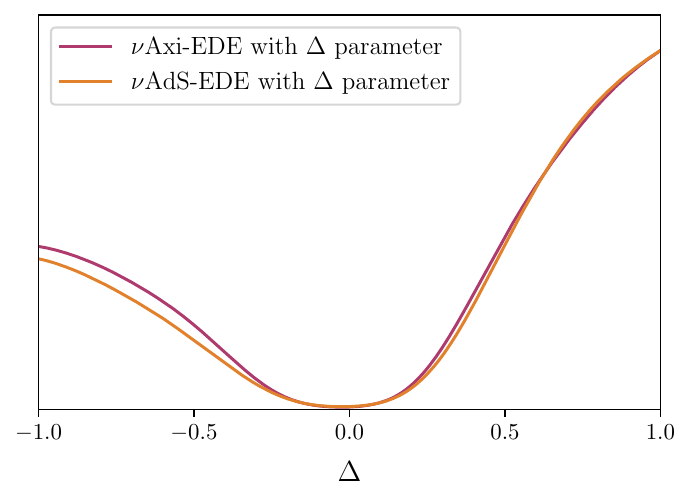}
    \caption{The 1D posterior distribution of $\Delta$ parameter in the $\nu$Axi-EDE and $\nu$AdS-EDE models. }
    \label{fig:delta}
\end{figure}

We use the \texttt{getdist} package \cite{Lewis:2019xzd} to plot the posterior distribution of cosmological parameters and analyse the best-fit parameters. The parameter constraints obtained from MCMC chains are summarized in \Table{tab:table1}. 
The combination dataset used for analysis is Planck+BAO+Pantheon+SH0ES. The corresponding best-fit values of $\chi^2$  are listed in the final line of the table. 
The one-dimensional posterior distribution of $m_{\min}$ and $\sum m_\nu$ for $\nu$Axi-EDE and $\nu$AdS-EDE models are shown in \Fig{fig:m}. In addition, \Fig{fig:delta} shows the posterior of $\Delta$ parameter.

\begin{table}[!htbp]
    \centering
    \scalebox{0.90}{
        \begin{tabular} { l c c c c c c}
        \toprule[1.0pt]
          & \multicolumn{3}{c}{$\nu$Axi-EDE} 
          & \multicolumn{3}{c}{$\nu$AdS-EDE} \\
        \cmidrule[0.3pt](lr{5pt}){2-4}\cmidrule[0.3pt](l{5pt}r){5-7}
         & base 
         & NH 
         & IH 
         & base 
         & NH 
         & IH\\
        \hline
        {$\log(10^{10} A_\mathrm{s})$}
        & $3.068\pm 0.015            $
        & $3.070^{+0.015}_{-0.016}   $
        & $3.075\pm 0.016            $
        & $3.068\pm 0.015            $
        & $3.070\pm 0.015            $
        & $3.075\pm 0.015            $
        \\
        
        {$n_\mathrm{s}   $} 
        & $0.9885\pm 0.0063          $
        & $0.9898\pm 0.0064          $
        & $0.9908\pm 0.0061          $
        & $0.9849\pm 0.0056          $
        & $0.9854\pm 0.0054          $
        & $0.9871\pm 0.0056          $
        \\
        
        {$\Omega_\mathrm{b} h^2$} 
        & $0.02284\pm 0.00021        $
        & $0.02285\pm 0.00022        $
        & $0.02285\pm 0.00021        $
        & $0.02303\pm 0.00020        $
        & $0.02305\pm 0.00020        $
        & $0.02310\pm 0.00020        $
        \\
        
        {$\Omega_\mathrm{c} h^2$} 
        & $0.1299\pm 0.0033          $
        & $0.1309^{+0.0032}_{-0.0036}$
        & $0.1315\pm 0.0035          $
        & $0.1284\pm 0.0030          $
        & $0.1286\pm 0.0030          $
        & $0.1293^{+0.0029}_{-0.0032}$
        \\
        
        {$\tau_\mathrm{reio}$}
        & $0.0583^{+0.0069}_{-0.0078}$
        & $0.0589^{+0.0072}_{-0.0082}$
        & $0.0606^{+0.0071}_{-0.0085}$
        & $0.0565\pm 0.0075          $
        & $0.0573\pm 0.0076          $
        & $0.0589^{+0.0068}_{-0.0078}$
        \\
        
        {$H_0\ [\kmsmpc]           $} 
        & $71.42^{+0.90}_{-0.79}     $
        & $71.51\pm 0.85             $
        & $71.52\pm 0.85             $
        & $71.04\pm 0.77             $
        & $70.98\pm 0.74             $
        & $70.95\pm 0.76             $
        \\
        
        $S_8                       $ 
        & $0.838\pm 0.012            $
        & $0.837\pm 0.013            $
        & $0.835\pm 0.013            $
        & $0.841\pm 0.013            $
        & $0.840\pm 0.013            $
        & $0.838\pm 0.013            $
        \\
        
        {$f_{\mathrm{EDE}}$} 
        & $0.109^{+0.028}_{-0.022}   $
        & $0.119\pm 0.025            $
        & $0.124^{+0.026}_{-0.023}   $
        & $0.073\pm 0.018            $
        & $0.087\pm 0.020            $
        & $0.093\pm 0.020            $
        \\
        
        {$\log_{10}a_c   $}
        & $-3.64^{+0.13}_{-0.20}     $
        & $-3.63^{+0.12}_{-0.012}    $
        & $-3.613^{+0.084}_{-0.014}  $
        & -
        & -
        & -
        \\
        
        {$\ln (1+z_c)    $} 
        & -
        & -
        & -
        & $8.181^{+0.095}_{-0.11}    $
        & $8.34\pm 0.10              $
        & $8.335\pm 0.094            $
        \\
        
        {$\alpha_{\mathrm{ads}}$} 
        & -
        & -
        & -
        & $< 0.000242                $
        & $< 0.000138                $
        & $< 0.000149                $
        \\
        
        {$m_{\min} \ [\ev]     $} 
        & -
        & $< 0.0429                  $
        & $< 0.0437                  $
        & -
        & $< 0.0363                  $
        & $< 0.0391                  $
        \\
        
        $\sum m_\nu \ [\ev]             $ 
        & -
        & $< 0.152                   $
        & $< 0.178                   $
        & -
        & $< 0.135                   $
        & $< 0.167                   $
        \\
        
        \hline
        $\chi^2                $ 
        & $3868.8$
        & $3863.7$
        & $3865.8$
        & $3864.9$
        & $3858.5$
        & $3861.2$
        \\
        \bottomrule[1.0pt]
        \end{tabular}
        }
    \caption{The mean values and $1\sigma$ constraints on the cosmological parameters in the base case, considering one massive neutrino, as well as the NH and IH cases of $\nu$Axi-EDE and $\nu$AdS-EDE models. The upper limits of $\alpha_{\mathrm{ads}}$, $m_{\min}$ and $\sum m_\nu$ are at 95\% C.L. }
    \label{tab:table1}
\end{table}




\textbf{Axion-like Early Dark Energy (Axi-EDE):} 
For the NH case of $\nu$Axi-EDE model, $m_{\min} < 0.0429$ eV and $\sum m_\nu < 0.152$ eV at 95\% C.L. While for the IH case, $m_{\min} < 0.0437\ev$ and $\sum m_\nu < 0.178 \ev$ at 95\% C.L.
Moreover, the constraints on other parameters for the NH or IH case of $\nu$Axi-EDE model exhibit negligible deviations from those in the base case. The NH and IH case of $\nu$Axi-EDE models show a better fit to the observational datasets, as evidenced by their smaller $\chi^2$ values when compared to the base case. Furthermore, the fitting performance of the NH case is better than the IH case, since $\Delta \chi^2 = \chi^2(\nu\text{Axi-EDE, NH}) - \chi^2(\nu\text{Axi-EDE, IH}) = -2.1$. 


The marginalized probability density $p(\Delta)$ for the NH and IH cases of $\nu$Axi-EDE model with parameter $\Delta$ are plotted in \Fig{fig:delta}. And then we can calculate these two probabilities
\begin{equation}
    P(\text{NH}) \equiv P(\Delta > 0) = \int_0^{1} p(\Delta) \mathrm{d}\Delta,
\end{equation}
and
\begin{equation}
    P(\text{IH}) \equiv P(\Delta > 0) = \int_{-1}^{0} p(\Delta) \mathrm{d}\Delta.
\end{equation}
The ratio of these two probabilities can be calculated as
\begin{equation}
    P(\mathrm{NH}):P(\mathrm{IH}) = 2.11:1 ,
\end{equation}
indicating that NH is more favorable than IH, which aligns with the aforementioned result of $\Delta \chi^2$ value.

\textbf{Early Dark Energy with AdS Phase (AdS-EDE):}
The constraints on the minimum neutrino mass in $\nu$AdS-EDE models are $m_{\min} < 0.0363\ev$ for the NH case, and $m_{\min} < 0.0391\ev$ for the IH case at 95\% C.L.
The constraints on the total neutrino mass are $\sum m_{\nu} < 0.135\ev$ for the NH case, and $\sum m_{\nu}< 0.167\ev$ for the IH case at 95\% C.L. 
The constraints on parameters other than neutrino parameters in the NH and IH cases of $\nu$AdS-EDE model are not significantly different from those in the base case. 
Moreover, the NH and IH cases of $\nu$AdS-EDE models fits better than the base case. 
The discrepancy between the $\chi^2$ values in NH and IH case is $\Delta\chi^2 = \chi^2(\nu\text{AdS-EDE, NH}) - \chi^2(\nu\text{AdS-EDE, IH}) = -2.7$. Therefore, the NH case still provides a better fit than the IH case within the framework of $\nu$AdS-EDE.

For the $\nu$AdS-EDE model with parameter $\Delta$, the ratio of $P(\mathrm{NH})$ and $P(\mathrm{IH})$
can be calculated from the $p(\Delta)$ plotted in \Fig{fig:delta},
\begin{equation}
    P(\mathrm{NH}):P(\mathrm{IH})=2.38:1 .
\end{equation}
It can be concluded that the NH is more favorable than the IH, which is  consistent with the aforementioned result of $\Delta \chi^2$ value.
Comparing the constraints of neutrino mass and the $\chi^2$ values of the above models, we find the results are congruent. The stringency of the neutrino mass constraints depends on the each specific model characteristics, but these constraints remain reliable for these two different early dark energy scenarios.


\section{Summary and Conclusions} \label{conclusions}
In this paper, we analyze the constraints on neutrino mass within consistent cosmological models, namely the $\nu$Axi-EDE and $\nu$AdS-EDE models. In both models, we assume three massive neutrino mass eigenstates with normal and inverted hierarchy. Additionally, we use the hierarchy parameter $\Delta$ in our analysis to evaluate the preference between the NH and IH. We employ a combination of \textit{Planck} + BAO + Pantheon + SH0ES measurements to constrain the parameters of these models. 

The resulting constraint on the minimum neutrino mass in $\nu$Axi-EDE model with NH is $m_{\min} < 0.0429$ $\ev$ at 95\% C.L., and the constraint on the sum of neutrino masses is $\sum m_\nu < 0.152$ $\ev$. For the $\nu$Axi-EDE model with IH, the neutrino mass constraints are $m_{\min}< 0.0437$ $\ev$ and $\sum m_\nu < 0.178$ $\ev$ both at 95\% C.L. 
%
The constraint on the minimum neutrino mass in the $\nu$AdS-EDE model with NH is $m_{\min}< 0.0363$ eV at 95\% C.L., and the constraint on the sum of neutrino masses is $\sum m_\nu < 0.135$ eV at 95\% C.L. In the case of $\nu$AdS-EDE with IH, the constraints are  $m_{\min}< 0.0391$ eV and $\sum m_\nu < 0.167$ eV at 95\% C.L. 

Compared to the $\nu$Axi-EDE base model with only one massive neutrino and a total neutrino mass of $\sum m_\nu = 0.06\ev$, there is no apparent shift in the constraints on cosmological parameters in the $\nu$Axi-EDE models with NH or IH.
The $\chi^2$ values of the $\nu$Axi-EDE model with NH and IH are both smaller than that of the base case. Additionally, we find that the $\nu$Axi-EDE model with NH fits better with the combination datasets compared to the other two cases. By integrating the probability distribution of $\Delta$ parameter, we find $P(\mathrm{NH}):P(\mathrm{IH}) = 2.11:1$ for $\nu$Axi-EDE, indicating a preference for the $\nu$Axi-EDE model with NH over IH.
Similarly, there seems to be no significant deviation of cosmological parameters in the $\nu$AdS-EDE model with NH or IH compared to those in the base case. The $\chi^2$ values of both NH and IH case in the $\nu$AdS-EDE model are smaller than those in the base case. 
Furthermore, $P(\mathrm{NH}):P(\mathrm{IH})=2.38:1$ indicates that the NH case of the $\nu$AdS-EDE model provides a better fit to the joint datasets compared to the IH case.

By comparing the overall results of the $\nu$Axi-EDE and $\nu$AdS-EDE models, it is found that different hierarchy of three massive neutrino eigenstates give similar results. Thus it is proved that the parameter constraints of these two early dark energy models are consistent and reliable.

\ 

\textit{Acknowledgements.}
We acknowledge the use of HPC Cluster of ITP-CAS. XZ is supported by grant from NSFC (Grant No. 12005183). QGH is supported by the grants from NSFC (Grant No.~12250010, 11975019, 11991052, 12047503), Key Research Program of Frontier Sciences, CAS, Grant No.~ZDBS-LY-7009.

\bibliography{refs.bib}

\begin{thebibliography}{61}%
\makeatletter
\providecommand \@ifxundefined [1]{%
 \@ifx{#1\undefined}
}%
\providecommand \@ifnum [1]{%
 \ifnum #1\expandafter \@firstoftwo
 \else \expandafter \@secondoftwo
 \fi
}%
\providecommand \@ifx [1]{%
 \ifx #1\expandafter \@firstoftwo
 \else \expandafter \@secondoftwo
 \fi
}%
\providecommand \natexlab [1]{#1}%
\providecommand \enquote  [1]{``#1''}%
\providecommand \bibnamefont  [1]{#1}%
\providecommand \bibfnamefont [1]{#1}%
\providecommand \citenamefont [1]{#1}%
\providecommand \href@noop [0]{\@secondoftwo}%
\providecommand \href [0]{\begingroup \@sanitize@url \@href}%
\providecommand \@href[1]{\@@startlink{#1}\@@href}%
\providecommand \@@href[1]{\endgroup#1\@@endlink}%
\providecommand \@sanitize@url [0]{\catcode `\\12\catcode `\$12\catcode
  `\&12\catcode `\#12\catcode `\^12\catcode `\_12\catcode `\%12\relax}%
\providecommand \@@startlink[1]{}%
\providecommand \@@endlink[0]{}%
\providecommand \url  [0]{\begingroup\@sanitize@url \@url }%
\providecommand \@url [1]{\endgroup\@href {#1}{\urlprefix }}%
\providecommand \urlprefix  [0]{URL }%
\providecommand \Eprint [0]{\href }%
\providecommand \doibase [0]{http://dx.doi.org/}%
\providecommand \selectlanguage [0]{\@gobble}%
\providecommand \bibinfo  [0]{\@secondoftwo}%
\providecommand \bibfield  [0]{\@secondoftwo}%
\providecommand \translation [1]{[#1]}%
\providecommand \BibitemOpen [0]{}%
\providecommand \bibitemStop [0]{}%
\providecommand \bibitemNoStop [0]{.\EOS\space}%
\providecommand \EOS [0]{\spacefactor3000\relax}%
\providecommand \BibitemShut  [1]{\csname bibitem#1\endcsname}%
\let\auto@bib@innerbib\@empty
\bibitem [{\citenamefont {Fukuda}\ \emph {et~al.}(1998)\citenamefont {Fukuda}
  \emph {et~al.}}]{Super-Kamiokande:1998kpq}%
  \BibitemOpen
  \bibfield  {author} {\bibinfo {author} {\bibfnamefont {Y.}~\bibnamefont
  {Fukuda}} \emph {et~al.} (\bibinfo {collaboration} {Super-Kamiokande}),\
  }\bibfield  {title} {\enquote {\bibinfo {title} {{Evidence for oscillation of
  atmospheric neutrinos}},}\ }\href {\doibase 10.1103/PhysRevLett.81.1562}
  {\bibfield  {journal} {\bibinfo  {journal} {Phys. Rev. Lett.}\ }\textbf
  {\bibinfo {volume} {81}},\ \bibinfo {pages} {1562--1567} (\bibinfo {year}
  {1998})},\ \Eprint {http://arxiv.org/abs/hep-ex/9807003}
  {arXiv:hep-ex/9807003} \BibitemShut {NoStop}%
\bibitem [{\citenamefont {Ahmad}\ \emph {et~al.}(2002)\citenamefont {Ahmad}
  \emph {et~al.}}]{SNO:2002tuh}%
  \BibitemOpen
  \bibfield  {author} {\bibinfo {author} {\bibfnamefont {Q.~R.}\ \bibnamefont
  {Ahmad}} \emph {et~al.} (\bibinfo {collaboration} {SNO}),\ }\bibfield
  {title} {\enquote {\bibinfo {title} {{Direct evidence for neutrino flavor
  transformation from neutral current interactions in the Sudbury Neutrino
  Observatory}},}\ }\href {\doibase 10.1103/PhysRevLett.89.011301} {\bibfield
  {journal} {\bibinfo  {journal} {Phys. Rev. Lett.}\ }\textbf {\bibinfo
  {volume} {89}},\ \bibinfo {pages} {011301} (\bibinfo {year} {2002})},\
  \Eprint {http://arxiv.org/abs/nucl-ex/0204008} {arXiv:nucl-ex/0204008}
  \BibitemShut {NoStop}%
\bibitem [{\citenamefont {Aker}\ \emph {et~al.}(2019)\citenamefont {Aker} \emph
  {et~al.}}]{KATRIN:2019yun}%
  \BibitemOpen
  \bibfield  {author} {\bibinfo {author} {\bibfnamefont {M.}~\bibnamefont
  {Aker}} \emph {et~al.} (\bibinfo {collaboration} {KATRIN}),\ }\bibfield
  {title} {\enquote {\bibinfo {title} {{Improved Upper Limit on the Neutrino
  Mass from a Direct Kinematic Method by KATRIN}},}\ }\href {\doibase
  10.1103/PhysRevLett.123.221802} {\bibfield  {journal} {\bibinfo  {journal}
  {Phys. Rev. Lett.}\ }\textbf {\bibinfo {volume} {123}},\ \bibinfo {pages}
  {221802} (\bibinfo {year} {2019})},\ \Eprint
  {http://arxiv.org/abs/1909.06048} {arXiv:1909.06048 [hep-ex]} \BibitemShut
  {NoStop}%
\bibitem [{\citenamefont {Olive}\ \emph {et~al.}(2014)\citenamefont {Olive}
  \emph {et~al.}}]{ParticleDataGroup:2014cgo}%
  \BibitemOpen
  \bibfield  {author} {\bibinfo {author} {\bibfnamefont {K.~A.}\ \bibnamefont
  {Olive}} \emph {et~al.} (\bibinfo {collaboration} {Particle Data Group}),\
  }\bibfield  {title} {\enquote {\bibinfo {title} {{Review of Particle
  Physics}},}\ }\href {\doibase 10.1088/1674-1137/38/9/090001} {\bibfield
  {journal} {\bibinfo  {journal} {Chin. Phys. C}\ }\textbf {\bibinfo {volume}
  {38}},\ \bibinfo {pages} {090001} (\bibinfo {year} {2014})}\BibitemShut
  {NoStop}%
\bibitem [{\citenamefont {Aghanim}\ \emph
  {et~al.}(2020{\natexlab{a}})\citenamefont {Aghanim} \emph
  {et~al.}}]{Planck:2018vyg}%
  \BibitemOpen
  \bibfield  {author} {\bibinfo {author} {\bibfnamefont {N.}~\bibnamefont
  {Aghanim}} \emph {et~al.} (\bibinfo {collaboration} {Planck}),\ }\bibfield
  {title} {\enquote {\bibinfo {title} {{Planck 2018 results. VI. Cosmological
  parameters}},}\ }\href {\doibase 10.1051/0004-6361/201833910} {\bibfield
  {journal} {\bibinfo  {journal} {Astron. Astrophys.}\ }\textbf {\bibinfo
  {volume} {641}},\ \bibinfo {pages} {A6} (\bibinfo {year}
  {2020}{\natexlab{a}})},\ \bibinfo {note} {[Erratum: Astron.Astrophys. 652, C4
  (2021)]},\ \Eprint {http://arxiv.org/abs/1807.06209} {arXiv:1807.06209
  [astro-ph.CO]} \BibitemShut {NoStop}%
\bibitem [{\citenamefont {Palanque-Delabrouille}\ \emph
  {et~al.}(2020)\citenamefont {Palanque-Delabrouille}, \citenamefont {Y\`eche},
  \citenamefont {Sch\"oneberg}, \citenamefont {Lesgourgues}, \citenamefont
  {Walther}, \citenamefont {Chabanier},\ and\ \citenamefont
  {Armengaud}}]{Palanque-Delabrouille:2019iyz}%
  \BibitemOpen
  \bibfield  {author} {\bibinfo {author} {\bibfnamefont {Nathalie}\
  \bibnamefont {Palanque-Delabrouille}}, \bibinfo {author} {\bibfnamefont
  {Christophe}\ \bibnamefont {Y\`eche}}, \bibinfo {author} {\bibfnamefont
  {Nils}\ \bibnamefont {Sch\"oneberg}}, \bibinfo {author} {\bibfnamefont
  {Julien}\ \bibnamefont {Lesgourgues}}, \bibinfo {author} {\bibfnamefont
  {Michael}\ \bibnamefont {Walther}}, \bibinfo {author} {\bibfnamefont
  {Sol\`ene}\ \bibnamefont {Chabanier}}, \ and\ \bibinfo {author}
  {\bibfnamefont {Eric}\ \bibnamefont {Armengaud}},\ }\bibfield  {title}
  {\enquote {\bibinfo {title} {{Hints, neutrino bounds and WDM constraints from
  SDSS DR14 Lyman-$\alpha$ and Planck full-survey data}},}\ }\href {\doibase
  10.1088/1475-7516/2020/04/038} {\bibfield  {journal} {\bibinfo  {journal}
  {JCAP}\ }\textbf {\bibinfo {volume} {04}},\ \bibinfo {pages} {038} (\bibinfo
  {year} {2020})},\ \Eprint {http://arxiv.org/abs/1911.09073} {arXiv:1911.09073
  [astro-ph.CO]} \BibitemShut {NoStop}%
\bibitem [{\citenamefont {Alam}\ \emph {et~al.}(2021)\citenamefont {Alam} \emph
  {et~al.}}]{eBOSS:2020yzd}%
  \BibitemOpen
  \bibfield  {author} {\bibinfo {author} {\bibfnamefont {Shadab}\ \bibnamefont
  {Alam}} \emph {et~al.} (\bibinfo {collaboration} {eBOSS}),\ }\bibfield
  {title} {\enquote {\bibinfo {title} {{Completed SDSS-IV extended Baryon
  Oscillation Spectroscopic Survey: Cosmological implications from two decades
  of spectroscopic surveys at the Apache Point Observatory}},}\ }\href
  {\doibase 10.1103/PhysRevD.103.083533} {\bibfield  {journal} {\bibinfo
  {journal} {Phys. Rev. D}\ }\textbf {\bibinfo {volume} {103}},\ \bibinfo
  {pages} {083533} (\bibinfo {year} {2021})},\ \Eprint
  {http://arxiv.org/abs/2007.08991} {arXiv:2007.08991 [astro-ph.CO]}
  \BibitemShut {NoStop}%
\bibitem [{\citenamefont {Abbott}\ \emph {et~al.}(2022)\citenamefont {Abbott}
  \emph {et~al.}}]{DES:2021wwk}%
  \BibitemOpen
  \bibfield  {author} {\bibinfo {author} {\bibfnamefont {T.~M.~C.}\
  \bibnamefont {Abbott}} \emph {et~al.} (\bibinfo {collaboration} {DES}),\
  }\bibfield  {title} {\enquote {\bibinfo {title} {{Dark Energy Survey Year 3
  results: Cosmological constraints from galaxy clustering and weak
  lensing}},}\ }\href {\doibase 10.1103/PhysRevD.105.023520} {\bibfield
  {journal} {\bibinfo  {journal} {Phys. Rev. D}\ }\textbf {\bibinfo {volume}
  {105}},\ \bibinfo {pages} {023520} (\bibinfo {year} {2022})},\ \Eprint
  {http://arxiv.org/abs/2105.13549} {arXiv:2105.13549 [astro-ph.CO]}
  \BibitemShut {NoStop}%
\bibitem [{\citenamefont {Wong}(2011)}]{Wong:2011ip}%
  \BibitemOpen
  \bibfield  {author} {\bibinfo {author} {\bibfnamefont {Yvonne Y.~Y.}\
  \bibnamefont {Wong}},\ }\bibfield  {title} {\enquote {\bibinfo {title}
  {{Neutrino mass in cosmology: status and prospects}},}\ }\href {\doibase
  10.1146/annurev-nucl-102010-130252} {\bibfield  {journal} {\bibinfo
  {journal} {Ann. Rev. Nucl. Part. Sci.}\ }\textbf {\bibinfo {volume} {61}},\
  \bibinfo {pages} {69--98} (\bibinfo {year} {2011})},\ \Eprint
  {http://arxiv.org/abs/1111.1436} {arXiv:1111.1436 [astro-ph.CO]} \BibitemShut
  {NoStop}%
\bibitem [{\citenamefont {Lesgourgues}\ and\ \citenamefont
  {Pastor}(2012)}]{Lesgourgues:2012uu}%
  \BibitemOpen
  \bibfield  {author} {\bibinfo {author} {\bibfnamefont {Julien}\ \bibnamefont
  {Lesgourgues}}\ and\ \bibinfo {author} {\bibfnamefont {Sergio}\ \bibnamefont
  {Pastor}},\ }\bibfield  {title} {\enquote {\bibinfo {title} {{Neutrino mass
  from Cosmology}},}\ }\href {\doibase 10.1155/2012/608515} {\bibfield
  {journal} {\bibinfo  {journal} {Adv. High Energy Phys.}\ }\textbf {\bibinfo
  {volume} {2012}},\ \bibinfo {pages} {608515} (\bibinfo {year} {2012})},\
  \Eprint {http://arxiv.org/abs/1212.6154} {arXiv:1212.6154 [hep-ph]}
  \BibitemShut {NoStop}%
\bibitem [{\citenamefont {Lesgourgues}\ \emph {et~al.}(2013)\citenamefont
  {Lesgourgues}, \citenamefont {Mangano}, \citenamefont {Miele},\ and\
  \citenamefont {Pastor}}]{Lesgourgues:2013sjj_Neutrino_Cosmology}%
  \BibitemOpen
  \bibfield  {author} {\bibinfo {author} {\bibfnamefont {Julien}\ \bibnamefont
  {Lesgourgues}}, \bibinfo {author} {\bibfnamefont {Gianpiero}\ \bibnamefont
  {Mangano}}, \bibinfo {author} {\bibfnamefont {Gennaro}\ \bibnamefont
  {Miele}}, \ and\ \bibinfo {author} {\bibfnamefont {Sergio}\ \bibnamefont
  {Pastor}},\ }\href@noop {} {\emph {\bibinfo {title} {{Neutrino Cosmology}}}}\
  (\bibinfo  {publisher} {Cambridge University Press},\ \bibinfo {year}
  {2013})\BibitemShut {NoStop}%
\bibitem [{\citenamefont {Lesgourgues}\ and\ \citenamefont
  {Pastor}(2014)}]{Lesgourgues:2014zoa}%
  \BibitemOpen
  \bibfield  {author} {\bibinfo {author} {\bibfnamefont {Julien}\ \bibnamefont
  {Lesgourgues}}\ and\ \bibinfo {author} {\bibfnamefont {Sergio}\ \bibnamefont
  {Pastor}},\ }\bibfield  {title} {\enquote {\bibinfo {title} {{Neutrino
  cosmology and Planck}},}\ }\href {\doibase 10.1088/1367-2630/16/6/065002}
  {\bibfield  {journal} {\bibinfo  {journal} {New J. Phys.}\ }\textbf {\bibinfo
  {volume} {16}},\ \bibinfo {pages} {065002} (\bibinfo {year} {2014})},\
  \Eprint {http://arxiv.org/abs/1404.1740} {arXiv:1404.1740 [hep-ph]}
  \BibitemShut {NoStop}%
\bibitem [{\citenamefont {Abazajian}\ \emph {et~al.}(2015)\citenamefont
  {Abazajian} \emph
  {et~al.}}]{TopicalConvenersKNAbazajianJECarlstromATLee:2013bxd}%
  \BibitemOpen
  \bibfield  {author} {\bibinfo {author} {\bibfnamefont {K.~N.}\ \bibnamefont
  {Abazajian}} \emph {et~al.} (\bibinfo {collaboration} {Topical Conveners:
  K.N. Abazajian, J.E. Carlstrom, A.T. Lee}),\ }\bibfield  {title} {\enquote
  {\bibinfo {title} {{Neutrino Physics from the Cosmic Microwave Background and
  Large Scale Structure}},}\ }\href {\doibase
  10.1016/j.astropartphys.2014.05.014} {\bibfield  {journal} {\bibinfo
  {journal} {Astropart. Phys.}\ }\textbf {\bibinfo {volume} {63}},\ \bibinfo
  {pages} {66--80} (\bibinfo {year} {2015})},\ \Eprint
  {http://arxiv.org/abs/1309.5383} {arXiv:1309.5383 [astro-ph.CO]} \BibitemShut
  {NoStop}%
\bibitem [{\citenamefont {Huang}\ \emph {et~al.}(2016)\citenamefont {Huang},
  \citenamefont {Wang},\ and\ \citenamefont {Wang}}]{Huang:2015wrx}%
  \BibitemOpen
  \bibfield  {author} {\bibinfo {author} {\bibfnamefont {Qing-Guo}\
  \bibnamefont {Huang}}, \bibinfo {author} {\bibfnamefont {Ke}~\bibnamefont
  {Wang}}, \ and\ \bibinfo {author} {\bibfnamefont {Sai}\ \bibnamefont
  {Wang}},\ }\bibfield  {title} {\enquote {\bibinfo {title} {{Constraints on
  the neutrino mass and mass hierarchy from cosmological observations}},}\
  }\href {\doibase 10.1140/epjc/s10052-016-4334-z} {\bibfield  {journal}
  {\bibinfo  {journal} {Eur. Phys. J. C}\ }\textbf {\bibinfo {volume} {76}},\
  \bibinfo {pages} {489} (\bibinfo {year} {2016})},\ \Eprint
  {http://arxiv.org/abs/1512.05899} {arXiv:1512.05899 [astro-ph.CO]}
  \BibitemShut {NoStop}%
\bibitem [{\citenamefont {Roy~Choudhury}\ and\ \citenamefont
  {Hannestad}(2020)}]{RoyChoudhury:2019hls}%
  \BibitemOpen
  \bibfield  {author} {\bibinfo {author} {\bibfnamefont {Shouvik}\ \bibnamefont
  {Roy~Choudhury}}\ and\ \bibinfo {author} {\bibfnamefont {Steen}\ \bibnamefont
  {Hannestad}},\ }\bibfield  {title} {\enquote {\bibinfo {title} {{Updated
  results on neutrino mass and mass hierarchy from cosmology with Planck 2018
  likelihoods}},}\ }\href {\doibase 10.1088/1475-7516/2020/07/037} {\bibfield
  {journal} {\bibinfo  {journal} {JCAP}\ }\textbf {\bibinfo {volume} {07}},\
  \bibinfo {pages} {037} (\bibinfo {year} {2020})},\ \Eprint
  {http://arxiv.org/abs/1907.12598} {arXiv:1907.12598 [astro-ph.CO]}
  \BibitemShut {NoStop}%
\bibitem [{\citenamefont {Reid}\ \emph {et~al.}(2010)\citenamefont {Reid},
  \citenamefont {Verde}, \citenamefont {Jimenez},\ and\ \citenamefont
  {Mena}}]{Reid:2009nq}%
  \BibitemOpen
  \bibfield  {author} {\bibinfo {author} {\bibfnamefont {Beth~A.}\ \bibnamefont
  {Reid}}, \bibinfo {author} {\bibfnamefont {Licia}\ \bibnamefont {Verde}},
  \bibinfo {author} {\bibfnamefont {Raul}\ \bibnamefont {Jimenez}}, \ and\
  \bibinfo {author} {\bibfnamefont {Olga}\ \bibnamefont {Mena}},\ }\bibfield
  {title} {\enquote {\bibinfo {title} {{Robust Neutrino Constraints by
  Combining Low Redshift Observations with the CMB}},}\ }\href {\doibase
  10.1088/1475-7516/2010/01/003} {\bibfield  {journal} {\bibinfo  {journal}
  {JCAP}\ }\textbf {\bibinfo {volume} {01}},\ \bibinfo {pages} {003} (\bibinfo
  {year} {2010})},\ \Eprint {http://arxiv.org/abs/0910.0008} {arXiv:0910.0008
  [astro-ph.CO]} \BibitemShut {NoStop}%
\bibitem [{\citenamefont {Kumar}\ and\ \citenamefont
  {Nunes}(2016)}]{Kumar:2016zpg}%
  \BibitemOpen
  \bibfield  {author} {\bibinfo {author} {\bibfnamefont {Suresh}\ \bibnamefont
  {Kumar}}\ and\ \bibinfo {author} {\bibfnamefont {Rafael~C.}\ \bibnamefont
  {Nunes}},\ }\bibfield  {title} {\enquote {\bibinfo {title} {{Probing the
  interaction between dark matter and dark energy in the presence of massive
  neutrinos}},}\ }\href {\doibase 10.1103/PhysRevD.94.123511} {\bibfield
  {journal} {\bibinfo  {journal} {Phys. Rev. D}\ }\textbf {\bibinfo {volume}
  {94}},\ \bibinfo {pages} {123511} (\bibinfo {year} {2016})},\ \Eprint
  {http://arxiv.org/abs/1608.02454} {arXiv:1608.02454 [astro-ph.CO]}
  \BibitemShut {NoStop}%
\bibitem [{\citenamefont {Yang}\ \emph {et~al.}(2017)\citenamefont {Yang},
  \citenamefont {Nunes}, \citenamefont {Pan},\ and\ \citenamefont
  {Mota}}]{Yang:2017amu}%
  \BibitemOpen
  \bibfield  {author} {\bibinfo {author} {\bibfnamefont {Weiqiang}\
  \bibnamefont {Yang}}, \bibinfo {author} {\bibfnamefont {Rafael~C.}\
  \bibnamefont {Nunes}}, \bibinfo {author} {\bibfnamefont {Supriya}\
  \bibnamefont {Pan}}, \ and\ \bibinfo {author} {\bibfnamefont {David~F.}\
  \bibnamefont {Mota}},\ }\bibfield  {title} {\enquote {\bibinfo {title}
  {{Effects of neutrino mass hierarchies on dynamical dark energy models}},}\
  }\href {\doibase 10.1103/PhysRevD.95.103522} {\bibfield  {journal} {\bibinfo
  {journal} {Phys. Rev. D}\ }\textbf {\bibinfo {volume} {95}},\ \bibinfo
  {pages} {103522} (\bibinfo {year} {2017})},\ \Eprint
  {http://arxiv.org/abs/1703.02556} {arXiv:1703.02556 [astro-ph.CO]}
  \BibitemShut {NoStop}%
\bibitem [{\citenamefont {Vagnozzi}\ \emph {et~al.}(2018)\citenamefont
  {Vagnozzi}, \citenamefont {Dhawan}, \citenamefont {Gerbino}, \citenamefont
  {Freese}, \citenamefont {Goobar},\ and\ \citenamefont
  {Mena}}]{Vagnozzi:2018jhn}%
  \BibitemOpen
  \bibfield  {author} {\bibinfo {author} {\bibfnamefont {Sunny}\ \bibnamefont
  {Vagnozzi}}, \bibinfo {author} {\bibfnamefont {Suhail}\ \bibnamefont
  {Dhawan}}, \bibinfo {author} {\bibfnamefont {Martina}\ \bibnamefont
  {Gerbino}}, \bibinfo {author} {\bibfnamefont {Katherine}\ \bibnamefont
  {Freese}}, \bibinfo {author} {\bibfnamefont {Ariel}\ \bibnamefont {Goobar}},
  \ and\ \bibinfo {author} {\bibfnamefont {Olga}\ \bibnamefont {Mena}},\
  }\bibfield  {title} {\enquote {\bibinfo {title} {{Constraints on the sum of
  the neutrino masses in dynamical dark energy models with $w(z) \geq -1$ are
  tighter than those obtained in $\Lambda$CDM}},}\ }\href {\doibase
  10.1103/PhysRevD.98.083501} {\bibfield  {journal} {\bibinfo  {journal} {Phys.
  Rev. D}\ }\textbf {\bibinfo {volume} {98}},\ \bibinfo {pages} {083501}
  (\bibinfo {year} {2018})},\ \Eprint {http://arxiv.org/abs/1801.08553}
  {arXiv:1801.08553 [astro-ph.CO]} \BibitemShut {NoStop}%
\bibitem [{\citenamefont {Guo}\ \emph {et~al.}(2018)\citenamefont {Guo},
  \citenamefont {Zhang},\ and\ \citenamefont {Zhang}}]{Guo:2018gyo}%
  \BibitemOpen
  \bibfield  {author} {\bibinfo {author} {\bibfnamefont {Rui-Yun}\ \bibnamefont
  {Guo}}, \bibinfo {author} {\bibfnamefont {Jing-Fei}\ \bibnamefont {Zhang}}, \
  and\ \bibinfo {author} {\bibfnamefont {Xin}\ \bibnamefont {Zhang}},\
  }\bibfield  {title} {\enquote {\bibinfo {title} {{Exploring neutrino mass and
  mass hierarchy in the scenario of vacuum energy interacting with cold dark
  matte}},}\ }\href {\doibase 10.1088/1674-1137/42/9/095103} {\bibfield
  {journal} {\bibinfo  {journal} {Chin. Phys. C}\ }\textbf {\bibinfo {volume}
  {42}},\ \bibinfo {pages} {095103} (\bibinfo {year} {2018})},\ \Eprint
  {http://arxiv.org/abs/1803.06910} {arXiv:1803.06910 [astro-ph.CO]}
  \BibitemShut {NoStop}%
\bibitem [{\citenamefont {Roy~Choudhury}\ and\ \citenamefont
  {Choubey}(2018)}]{RoyChoudhury:2018gay}%
  \BibitemOpen
  \bibfield  {author} {\bibinfo {author} {\bibfnamefont {Shouvik}\ \bibnamefont
  {Roy~Choudhury}}\ and\ \bibinfo {author} {\bibfnamefont {Sandhya}\
  \bibnamefont {Choubey}},\ }\bibfield  {title} {\enquote {\bibinfo {title}
  {{Updated Bounds on Sum of Neutrino Masses in Various Cosmological
  Scenarios}},}\ }\href {\doibase 10.1088/1475-7516/2018/09/017} {\bibfield
  {journal} {\bibinfo  {journal} {JCAP}\ }\textbf {\bibinfo {volume} {09}},\
  \bibinfo {pages} {017} (\bibinfo {year} {2018})},\ \Eprint
  {http://arxiv.org/abs/1806.10832} {arXiv:1806.10832 [astro-ph.CO]}
  \BibitemShut {NoStop}%
\bibitem [{\citenamefont {Feng}\ \emph
  {et~al.}(2020{\natexlab{a}})\citenamefont {Feng}, \citenamefont {Li},
  \citenamefont {Zhang},\ and\ \citenamefont {Zhang}}]{Feng:2019mym}%
  \BibitemOpen
  \bibfield  {author} {\bibinfo {author} {\bibfnamefont {Lu}~\bibnamefont
  {Feng}}, \bibinfo {author} {\bibfnamefont {Hai-Li}\ \bibnamefont {Li}},
  \bibinfo {author} {\bibfnamefont {Jing-Fei}\ \bibnamefont {Zhang}}, \ and\
  \bibinfo {author} {\bibfnamefont {Xin}\ \bibnamefont {Zhang}},\ }\bibfield
  {title} {\enquote {\bibinfo {title} {{Exploring neutrino mass and mass
  hierarchy in interacting dark energy models}},}\ }\href {\doibase
  10.1007/s11433-019-9431-9} {\bibfield  {journal} {\bibinfo  {journal} {Sci.
  China Phys. Mech. Astron.}\ }\textbf {\bibinfo {volume} {63}},\ \bibinfo
  {pages} {220401} (\bibinfo {year} {2020}{\natexlab{a}})},\ \Eprint
  {http://arxiv.org/abs/1903.08848} {arXiv:1903.08848 [astro-ph.CO]}
  \BibitemShut {NoStop}%
\bibitem [{\citenamefont {Feng}\ \emph
  {et~al.}(2020{\natexlab{b}})\citenamefont {Feng}, \citenamefont {He},
  \citenamefont {Li}, \citenamefont {Zhang},\ and\ \citenamefont
  {Zhang}}]{Feng:2019jqa}%
  \BibitemOpen
  \bibfield  {author} {\bibinfo {author} {\bibfnamefont {Lu}~\bibnamefont
  {Feng}}, \bibinfo {author} {\bibfnamefont {Dong-Ze}\ \bibnamefont {He}},
  \bibinfo {author} {\bibfnamefont {Hai-Li}\ \bibnamefont {Li}}, \bibinfo
  {author} {\bibfnamefont {Jing-Fei}\ \bibnamefont {Zhang}}, \ and\ \bibinfo
  {author} {\bibfnamefont {Xin}\ \bibnamefont {Zhang}},\ }\bibfield  {title}
  {\enquote {\bibinfo {title} {{Constraints on active and sterile neutrinos in
  an interacting dark energy cosmology}},}\ }\href {\doibase
  10.1007/s11433-019-1511-8} {\bibfield  {journal} {\bibinfo  {journal} {Sci.
  China Phys. Mech. Astron.}\ }\textbf {\bibinfo {volume} {63}},\ \bibinfo
  {pages} {290404} (\bibinfo {year} {2020}{\natexlab{b}})},\ \Eprint
  {http://arxiv.org/abs/1910.03872} {arXiv:1910.03872 [astro-ph.CO]}
  \BibitemShut {NoStop}%
\bibitem [{\citenamefont {Li}\ \emph {et~al.}(2020)\citenamefont {Li},
  \citenamefont {Zhang},\ and\ \citenamefont {Zhang}}]{Li:2020gtk}%
  \BibitemOpen
  \bibfield  {author} {\bibinfo {author} {\bibfnamefont {Hai-Li}\ \bibnamefont
  {Li}}, \bibinfo {author} {\bibfnamefont {Jing-Fei}\ \bibnamefont {Zhang}}, \
  and\ \bibinfo {author} {\bibfnamefont {Xin}\ \bibnamefont {Zhang}},\
  }\bibfield  {title} {\enquote {\bibinfo {title} {{Constraints on neutrino
  mass in the scenario of vacuum energy interacting with cold dark matter after
  Planck 2018}},}\ }\href {\doibase 10.1088/1572-9494/abb7c9} {\bibfield
  {journal} {\bibinfo  {journal} {Commun. Theor. Phys.}\ }\textbf {\bibinfo
  {volume} {72}},\ \bibinfo {pages} {125401} (\bibinfo {year} {2020})},\
  \Eprint {http://arxiv.org/abs/2005.12041} {arXiv:2005.12041 [astro-ph.CO]}
  \BibitemShut {NoStop}%
\bibitem [{\citenamefont {Yang}\ \emph {et~al.}(2021)\citenamefont {Yang},
  \citenamefont {Di~Valentino}, \citenamefont {Pan},\ and\ \citenamefont
  {Mena}}]{Yang:2020ope}%
  \BibitemOpen
  \bibfield  {author} {\bibinfo {author} {\bibfnamefont {Weiqiang}\
  \bibnamefont {Yang}}, \bibinfo {author} {\bibfnamefont {Eleonora}\
  \bibnamefont {Di~Valentino}}, \bibinfo {author} {\bibfnamefont {Supriya}\
  \bibnamefont {Pan}}, \ and\ \bibinfo {author} {\bibfnamefont {Olga}\
  \bibnamefont {Mena}},\ }\bibfield  {title} {\enquote {\bibinfo {title}
  {{Emergent Dark Energy, neutrinos and cosmological tensions}},}\ }\href
  {\doibase 10.1016/j.dark.2020.100762} {\bibfield  {journal} {\bibinfo
  {journal} {Phys. Dark Univ.}\ }\textbf {\bibinfo {volume} {31}},\ \bibinfo
  {pages} {100762} (\bibinfo {year} {2021})},\ \Eprint
  {http://arxiv.org/abs/2007.02927} {arXiv:2007.02927 [astro-ph.CO]}
  \BibitemShut {NoStop}%
\bibitem [{\citenamefont {Cardona}\ \emph {et~al.}(2021)\citenamefont
  {Cardona}, \citenamefont {Arjona}, \citenamefont {Estrada},\ and\
  \citenamefont {Nesseris}}]{Cardona:2020ama}%
  \BibitemOpen
  \bibfield  {author} {\bibinfo {author} {\bibfnamefont {Wilmar}\ \bibnamefont
  {Cardona}}, \bibinfo {author} {\bibfnamefont {Rub\'en}\ \bibnamefont
  {Arjona}}, \bibinfo {author} {\bibfnamefont {Alejandro}\ \bibnamefont
  {Estrada}}, \ and\ \bibinfo {author} {\bibfnamefont {Savvas}\ \bibnamefont
  {Nesseris}},\ }\bibfield  {title} {\enquote {\bibinfo {title} {{Cosmological
  constraints with the Effective Fluid approach for Modified Gravity}},}\
  }\href {\doibase 10.1088/1475-7516/2021/05/064} {\bibfield  {journal}
  {\bibinfo  {journal} {JCAP}\ }\textbf {\bibinfo {volume} {05}},\ \bibinfo
  {pages} {064} (\bibinfo {year} {2021})},\ \Eprint
  {http://arxiv.org/abs/2012.05282} {arXiv:2012.05282 [astro-ph.CO]}
  \BibitemShut {NoStop}%
\bibitem [{\citenamefont {G\'omez-Valent}\ \emph {et~al.}(2022)\citenamefont
  {G\'omez-Valent}, \citenamefont {Zheng}, \citenamefont {Amendola},
  \citenamefont {Wetterich},\ and\ \citenamefont
  {Pettorino}}]{Gomez-Valent:2022bku}%
  \BibitemOpen
  \bibfield  {author} {\bibinfo {author} {\bibfnamefont {Adri\`a}\ \bibnamefont
  {G\'omez-Valent}}, \bibinfo {author} {\bibfnamefont {Ziyang}\ \bibnamefont
  {Zheng}}, \bibinfo {author} {\bibfnamefont {Luca}\ \bibnamefont {Amendola}},
  \bibinfo {author} {\bibfnamefont {Christof}\ \bibnamefont {Wetterich}}, \
  and\ \bibinfo {author} {\bibfnamefont {Valeria}\ \bibnamefont {Pettorino}},\
  }\bibfield  {title} {\enquote {\bibinfo {title} {{Coupled and uncoupled early
  dark energy, massive neutrinos, and the cosmological tensions}},}\ }\href
  {\doibase 10.1103/PhysRevD.106.103522} {\bibfield  {journal} {\bibinfo
  {journal} {Phys. Rev. D}\ }\textbf {\bibinfo {volume} {106}},\ \bibinfo
  {pages} {103522} (\bibinfo {year} {2022})},\ \Eprint
  {http://arxiv.org/abs/2207.14487} {arXiv:2207.14487 [astro-ph.CO]}
  \BibitemShut {NoStop}%
\bibitem [{\citenamefont {Yadav}\ \emph {et~al.}(2023)\citenamefont {Yadav},
  \citenamefont {Yadav},\ and\ \citenamefont {Yadav}}]{Yadav:2023qfj}%
  \BibitemOpen
  \bibfield  {author} {\bibinfo {author} {\bibfnamefont {Vikrant}\ \bibnamefont
  {Yadav}}, \bibinfo {author} {\bibfnamefont {Santosh~Kumar}\ \bibnamefont
  {Yadav}}, \ and\ \bibinfo {author} {\bibfnamefont {Anil~Kumar}\ \bibnamefont
  {Yadav}},\ }\bibfield  {title} {\enquote {\bibinfo {title} {{Observational
  Constraints on generalized dark matter properties in the presence of
  neutrinos with the final Planck release}},}\ }\href {\doibase
  10.1016/j.dark.2023.101363} {\bibfield  {journal} {\bibinfo  {journal} {Phys.
  Dark Univ.}\ }\textbf {\bibinfo {volume} {42}},\ \bibinfo {pages} {101363}
  (\bibinfo {year} {2023})},\ \Eprint {http://arxiv.org/abs/2307.05155}
  {arXiv:2307.05155 [astro-ph.CO]} \BibitemShut {NoStop}%
\bibitem [{\citenamefont {Lu}\ \emph {et~al.}(2016)\citenamefont {Lu},
  \citenamefont {Liu}, \citenamefont {Wu}, \citenamefont {Wang},\ and\
  \citenamefont {Yang}}]{Lu:2016hsd}%
  \BibitemOpen
  \bibfield  {author} {\bibinfo {author} {\bibfnamefont {Jianbo}\ \bibnamefont
  {Lu}}, \bibinfo {author} {\bibfnamefont {Molin}\ \bibnamefont {Liu}},
  \bibinfo {author} {\bibfnamefont {Yabo}\ \bibnamefont {Wu}}, \bibinfo
  {author} {\bibfnamefont {Yan}\ \bibnamefont {Wang}}, \ and\ \bibinfo {author}
  {\bibfnamefont {Weiqiang}\ \bibnamefont {Yang}},\ }\bibfield  {title}
  {\enquote {\bibinfo {title} {{Cosmic constraint on massive neutrinos in
  viable $f(R)$ gravity with producing $\Lambda$CDM background expansion}},}\
  }\href {\doibase 10.1140/epjc/s10052-016-4525-7} {\bibfield  {journal}
  {\bibinfo  {journal} {Eur. Phys. J. C}\ }\textbf {\bibinfo {volume} {76}},\
  \bibinfo {pages} {679} (\bibinfo {year} {2016})},\ \Eprint
  {http://arxiv.org/abs/1606.02987} {arXiv:1606.02987 [astro-ph.CO]}
  \BibitemShut {NoStop}%
\bibitem [{\citenamefont {Sekiguchi}\ and\ \citenamefont
  {Takahashi}(2021)}]{Sekiguchi:2020igz}%
  \BibitemOpen
  \bibfield  {author} {\bibinfo {author} {\bibfnamefont {Toyokazu}\
  \bibnamefont {Sekiguchi}}\ and\ \bibinfo {author} {\bibfnamefont {Tomo}\
  \bibnamefont {Takahashi}},\ }\bibfield  {title} {\enquote {\bibinfo {title}
  {{Cosmological bound on neutrino masses in the light of $H_0$ tension}},}\
  }\href {\doibase 10.1103/PhysRevD.103.083516} {\bibfield  {journal} {\bibinfo
   {journal} {Phys. Rev. D}\ }\textbf {\bibinfo {volume} {103}},\ \bibinfo
  {pages} {083516} (\bibinfo {year} {2021})},\ \Eprint
  {http://arxiv.org/abs/2011.14481} {arXiv:2011.14481 [astro-ph.CO]}
  \BibitemShut {NoStop}%
\bibitem [{\citenamefont {Di~Valentino}\ \emph {et~al.}(2021)\citenamefont
  {Di~Valentino}, \citenamefont {Mena}, \citenamefont {Pan}, \citenamefont
  {Visinelli}, \citenamefont {Yang}, \citenamefont {Melchiorri}, \citenamefont
  {Mota}, \citenamefont {Riess},\ and\ \citenamefont
  {Silk}}]{DiValentino:2021izs}%
  \BibitemOpen
  \bibfield  {author} {\bibinfo {author} {\bibfnamefont {Eleonora}\
  \bibnamefont {Di~Valentino}}, \bibinfo {author} {\bibfnamefont {Olga}\
  \bibnamefont {Mena}}, \bibinfo {author} {\bibfnamefont {Supriya}\
  \bibnamefont {Pan}}, \bibinfo {author} {\bibfnamefont {Luca}\ \bibnamefont
  {Visinelli}}, \bibinfo {author} {\bibfnamefont {Weiqiang}\ \bibnamefont
  {Yang}}, \bibinfo {author} {\bibfnamefont {Alessandro}\ \bibnamefont
  {Melchiorri}}, \bibinfo {author} {\bibfnamefont {David~F.}\ \bibnamefont
  {Mota}}, \bibinfo {author} {\bibfnamefont {Adam~G.}\ \bibnamefont {Riess}}, \
  and\ \bibinfo {author} {\bibfnamefont {Joseph}\ \bibnamefont {Silk}},\
  }\bibfield  {title} {\enquote {\bibinfo {title} {{In the realm of the Hubble
  tension\textemdash{}a review of solutions}},}\ }\href {\doibase
  10.1088/1361-6382/ac086d} {\bibfield  {journal} {\bibinfo  {journal} {Class.
  Quant. Grav.}\ }\textbf {\bibinfo {volume} {38}},\ \bibinfo {pages} {153001}
  (\bibinfo {year} {2021})},\ \Eprint {http://arxiv.org/abs/2103.01183}
  {arXiv:2103.01183 [astro-ph.CO]} \BibitemShut {NoStop}%
\bibitem [{\citenamefont {Abdalla}\ \emph {et~al.}(2022)\citenamefont {Abdalla}
  \emph {et~al.}}]{Abdalla:2022yfr}%
  \BibitemOpen
  \bibfield  {author} {\bibinfo {author} {\bibfnamefont {Elcio}\ \bibnamefont
  {Abdalla}} \emph {et~al.},\ }\bibfield  {title} {\enquote {\bibinfo {title}
  {{Cosmology intertwined: A review of the particle physics, astrophysics, and
  cosmology associated with the cosmological tensions and anomalies}},}\ }\href
  {\doibase 10.1016/j.jheap.2022.04.002} {\bibfield  {journal} {\bibinfo
  {journal} {JHEAp}\ }\textbf {\bibinfo {volume} {34}},\ \bibinfo {pages}
  {49--211} (\bibinfo {year} {2022})},\ \Eprint
  {http://arxiv.org/abs/2203.06142} {arXiv:2203.06142 [astro-ph.CO]}
  \BibitemShut {NoStop}%
\bibitem [{\citenamefont {Aghanim}\ \emph
  {et~al.}(2020{\natexlab{b}})\citenamefont {Aghanim} \emph
  {et~al.}}]{Planck:2018nkj}%
  \BibitemOpen
  \bibfield  {author} {\bibinfo {author} {\bibfnamefont {N.}~\bibnamefont
  {Aghanim}} \emph {et~al.} (\bibinfo {collaboration} {Planck}),\ }\bibfield
  {title} {\enquote {\bibinfo {title} {{Planck 2018 results. I. Overview and
  the cosmological legacy of Planck}},}\ }\href {\doibase
  10.1051/0004-6361/201833880} {\bibfield  {journal} {\bibinfo  {journal}
  {Astron. Astrophys.}\ }\textbf {\bibinfo {volume} {641}},\ \bibinfo {pages}
  {A1} (\bibinfo {year} {2020}{\natexlab{b}})},\ \Eprint
  {http://arxiv.org/abs/1807.06205} {arXiv:1807.06205 [astro-ph.CO]}
  \BibitemShut {NoStop}%
\bibitem [{\citenamefont {Riess}\ \emph {et~al.}(2022)\citenamefont {Riess}
  \emph {et~al.}}]{Riess:2021jrx}%
  \BibitemOpen
  \bibfield  {author} {\bibinfo {author} {\bibfnamefont {Adam~G.}\ \bibnamefont
  {Riess}} \emph {et~al.},\ }\bibfield  {title} {\enquote {\bibinfo {title} {{A
  Comprehensive Measurement of the Local Value of the Hubble Constant with 1 km
  s$^{-1}$ Mpc$^{-1}$ Uncertainty from the Hubble Space Telescope and the SH0ES
  Team}},}\ }\href {\doibase 10.3847/2041-8213/ac5c5b} {\bibfield  {journal}
  {\bibinfo  {journal} {Astrophys. J. Lett.}\ }\textbf {\bibinfo {volume}
  {934}},\ \bibinfo {pages} {L7} (\bibinfo {year} {2022})},\ \Eprint
  {http://arxiv.org/abs/2112.04510} {arXiv:2112.04510 [astro-ph.CO]}
  \BibitemShut {NoStop}%
\bibitem [{\citenamefont {Poulin}\ \emph {et~al.}(2023)\citenamefont {Poulin},
  \citenamefont {Smith},\ and\ \citenamefont {Karwal}}]{Poulin:2023lkg}%
  \BibitemOpen
  \bibfield  {author} {\bibinfo {author} {\bibfnamefont {Vivian}\ \bibnamefont
  {Poulin}}, \bibinfo {author} {\bibfnamefont {Tristan~L.}\ \bibnamefont
  {Smith}}, \ and\ \bibinfo {author} {\bibfnamefont {Tanvi}\ \bibnamefont
  {Karwal}},\ }\bibfield  {title} {\enquote {\bibinfo {title} {{The Ups and
  Downs of Early Dark Energy solutions to the Hubble tension: a review of
  models, hints and constraints circa 2023}},}\ }\href@noop {} {\  (\bibinfo
  {year} {2023})},\ \Eprint {http://arxiv.org/abs/2302.09032} {arXiv:2302.09032
  [astro-ph.CO]} \BibitemShut {NoStop}%
\bibitem [{\citenamefont {Poulin}\ \emph {et~al.}(2019)\citenamefont {Poulin},
  \citenamefont {Smith}, \citenamefont {Karwal},\ and\ \citenamefont
  {Kamionkowski}}]{Poulin:2018cxd}%
  \BibitemOpen
  \bibfield  {author} {\bibinfo {author} {\bibfnamefont {Vivian}\ \bibnamefont
  {Poulin}}, \bibinfo {author} {\bibfnamefont {Tristan~L.}\ \bibnamefont
  {Smith}}, \bibinfo {author} {\bibfnamefont {Tanvi}\ \bibnamefont {Karwal}}, \
  and\ \bibinfo {author} {\bibfnamefont {Marc}\ \bibnamefont {Kamionkowski}},\
  }\bibfield  {title} {\enquote {\bibinfo {title} {{Early Dark Energy Can
  Resolve The Hubble Tension}},}\ }\href {\doibase
  10.1103/PhysRevLett.122.221301} {\bibfield  {journal} {\bibinfo  {journal}
  {Phys. Rev. Lett.}\ }\textbf {\bibinfo {volume} {122}},\ \bibinfo {pages}
  {221301} (\bibinfo {year} {2019})},\ \Eprint
  {http://arxiv.org/abs/1811.04083} {arXiv:1811.04083 [astro-ph.CO]}
  \BibitemShut {NoStop}%
\bibitem [{\citenamefont {Lin}\ \emph {et~al.}(2019)\citenamefont {Lin},
  \citenamefont {Benevento}, \citenamefont {Hu},\ and\ \citenamefont
  {Raveri}}]{Lin:2019qug}%
  \BibitemOpen
  \bibfield  {author} {\bibinfo {author} {\bibfnamefont {Meng-Xiang}\
  \bibnamefont {Lin}}, \bibinfo {author} {\bibfnamefont {Giampaolo}\
  \bibnamefont {Benevento}}, \bibinfo {author} {\bibfnamefont {Wayne}\
  \bibnamefont {Hu}}, \ and\ \bibinfo {author} {\bibfnamefont {Marco}\
  \bibnamefont {Raveri}},\ }\bibfield  {title} {\enquote {\bibinfo {title}
  {{Acoustic Dark Energy: Potential Conversion of the Hubble Tension}},}\
  }\href {\doibase 10.1103/PhysRevD.100.063542} {\bibfield  {journal} {\bibinfo
   {journal} {Phys. Rev. D}\ }\textbf {\bibinfo {volume} {100}},\ \bibinfo
  {pages} {063542} (\bibinfo {year} {2019})},\ \Eprint
  {http://arxiv.org/abs/1905.12618} {arXiv:1905.12618 [astro-ph.CO]}
  \BibitemShut {NoStop}%
\bibitem [{\citenamefont {Smith}\ \emph {et~al.}(2020)\citenamefont {Smith},
  \citenamefont {Poulin},\ and\ \citenamefont {Amin}}]{Smith:2019ihp}%
  \BibitemOpen
  \bibfield  {author} {\bibinfo {author} {\bibfnamefont {Tristan~L.}\
  \bibnamefont {Smith}}, \bibinfo {author} {\bibfnamefont {Vivian}\
  \bibnamefont {Poulin}}, \ and\ \bibinfo {author} {\bibfnamefont {Mustafa~A.}\
  \bibnamefont {Amin}},\ }\bibfield  {title} {\enquote {\bibinfo {title}
  {{Oscillating scalar fields and the Hubble tension: a resolution with novel
  signatures}},}\ }\href {\doibase 10.1103/PhysRevD.101.063523} {\bibfield
  {journal} {\bibinfo  {journal} {Phys. Rev. D}\ }\textbf {\bibinfo {volume}
  {101}},\ \bibinfo {pages} {063523} (\bibinfo {year} {2020})},\ \Eprint
  {http://arxiv.org/abs/1908.06995} {arXiv:1908.06995 [astro-ph.CO]}
  \BibitemShut {NoStop}%
\bibitem [{\citenamefont {Agrawal}\ \emph {et~al.}(2023)\citenamefont
  {Agrawal}, \citenamefont {Cyr-Racine}, \citenamefont {Pinner},\ and\
  \citenamefont {Randall}}]{Agrawal:2019lmo}%
  \BibitemOpen
  \bibfield  {author} {\bibinfo {author} {\bibfnamefont {Prateek}\ \bibnamefont
  {Agrawal}}, \bibinfo {author} {\bibfnamefont {Francis-Yan}\ \bibnamefont
  {Cyr-Racine}}, \bibinfo {author} {\bibfnamefont {David}\ \bibnamefont
  {Pinner}}, \ and\ \bibinfo {author} {\bibfnamefont {Lisa}\ \bibnamefont
  {Randall}},\ }\bibfield  {title} {\enquote {\bibinfo {title} {{Rock
  \textquoteleft{}n\textquoteright{} roll solutions to the Hubble tension}},}\
  }\href {\doibase 10.1016/j.dark.2023.101347} {\bibfield  {journal} {\bibinfo
  {journal} {Phys. Dark Univ.}\ }\textbf {\bibinfo {volume} {42}},\ \bibinfo
  {pages} {101347} (\bibinfo {year} {2023})},\ \Eprint
  {http://arxiv.org/abs/1904.01016} {arXiv:1904.01016 [astro-ph.CO]}
  \BibitemShut {NoStop}%
\bibitem [{\citenamefont {Braglia}\ \emph {et~al.}(2020)\citenamefont
  {Braglia}, \citenamefont {Emond}, \citenamefont {Finelli}, \citenamefont
  {Gumrukcuoglu},\ and\ \citenamefont {Koyama}}]{Braglia:2020bym}%
  \BibitemOpen
  \bibfield  {author} {\bibinfo {author} {\bibfnamefont {Matteo}\ \bibnamefont
  {Braglia}}, \bibinfo {author} {\bibfnamefont {William~T.}\ \bibnamefont
  {Emond}}, \bibinfo {author} {\bibfnamefont {Fabio}\ \bibnamefont {Finelli}},
  \bibinfo {author} {\bibfnamefont {A.~Emir}\ \bibnamefont {Gumrukcuoglu}}, \
  and\ \bibinfo {author} {\bibfnamefont {Kazuya}\ \bibnamefont {Koyama}},\
  }\bibfield  {title} {\enquote {\bibinfo {title} {{Unified framework for early
  dark energy from $\alpha$-attractors}},}\ }\href {\doibase
  10.1103/PhysRevD.102.083513} {\bibfield  {journal} {\bibinfo  {journal}
  {Phys. Rev. D}\ }\textbf {\bibinfo {volume} {102}},\ \bibinfo {pages}
  {083513} (\bibinfo {year} {2020})},\ \Eprint
  {http://arxiv.org/abs/2005.14053} {arXiv:2005.14053 [astro-ph.CO]}
  \BibitemShut {NoStop}%
\bibitem [{\citenamefont {Niedermann}\ and\ \citenamefont
  {Sloth}(2021)}]{Niedermann:2019olb}%
  \BibitemOpen
  \bibfield  {author} {\bibinfo {author} {\bibfnamefont {Florian}\ \bibnamefont
  {Niedermann}}\ and\ \bibinfo {author} {\bibfnamefont {Martin~S.}\
  \bibnamefont {Sloth}},\ }\bibfield  {title} {\enquote {\bibinfo {title} {{New
  early dark energy}},}\ }\href {\doibase 10.1103/PhysRevD.103.L041303}
  {\bibfield  {journal} {\bibinfo  {journal} {Phys. Rev. D}\ }\textbf {\bibinfo
  {volume} {103}},\ \bibinfo {pages} {L041303} (\bibinfo {year} {2021})},\
  \Eprint {http://arxiv.org/abs/1910.10739} {arXiv:1910.10739 [astro-ph.CO]}
  \BibitemShut {NoStop}%
\bibitem [{\citenamefont {Ye}\ and\ \citenamefont {Piao}(2020)}]{Ye:2020btb}%
  \BibitemOpen
  \bibfield  {author} {\bibinfo {author} {\bibfnamefont {Gen}\ \bibnamefont
  {Ye}}\ and\ \bibinfo {author} {\bibfnamefont {Yun-Song}\ \bibnamefont
  {Piao}},\ }\bibfield  {title} {\enquote {\bibinfo {title} {{Is the Hubble
  tension a hint of AdS phase around recombination?}}}\ }\href {\doibase
  10.1103/PhysRevD.101.083507} {\bibfield  {journal} {\bibinfo  {journal}
  {Phys. Rev. D}\ }\textbf {\bibinfo {volume} {101}},\ \bibinfo {pages}
  {083507} (\bibinfo {year} {2020})},\ \Eprint
  {http://arxiv.org/abs/2001.02451} {arXiv:2001.02451 [astro-ph.CO]}
  \BibitemShut {NoStop}%
\bibitem [{\citenamefont {Niedermann}\ and\ \citenamefont
  {Sloth}(2020)}]{Niedermann:2020dwg}%
  \BibitemOpen
  \bibfield  {author} {\bibinfo {author} {\bibfnamefont {Florian}\ \bibnamefont
  {Niedermann}}\ and\ \bibinfo {author} {\bibfnamefont {Martin~S.}\
  \bibnamefont {Sloth}},\ }\bibfield  {title} {\enquote {\bibinfo {title}
  {{Resolving the Hubble tension with new early dark energy}},}\ }\href
  {\doibase 10.1103/PhysRevD.102.063527} {\bibfield  {journal} {\bibinfo
  {journal} {Phys. Rev. D}\ }\textbf {\bibinfo {volume} {102}},\ \bibinfo
  {pages} {063527} (\bibinfo {year} {2020})},\ \Eprint
  {http://arxiv.org/abs/2006.06686} {arXiv:2006.06686 [astro-ph.CO]}
  \BibitemShut {NoStop}%
\bibitem [{\citenamefont {Karwal}\ \emph {et~al.}(2022)\citenamefont {Karwal},
  \citenamefont {Raveri}, \citenamefont {Jain}, \citenamefont {Khoury},\ and\
  \citenamefont {Trodden}}]{Karwal:2021vpk}%
  \BibitemOpen
  \bibfield  {author} {\bibinfo {author} {\bibfnamefont {Tanvi}\ \bibnamefont
  {Karwal}}, \bibinfo {author} {\bibfnamefont {Marco}\ \bibnamefont {Raveri}},
  \bibinfo {author} {\bibfnamefont {Bhuvnesh}\ \bibnamefont {Jain}}, \bibinfo
  {author} {\bibfnamefont {Justin}\ \bibnamefont {Khoury}}, \ and\ \bibinfo
  {author} {\bibfnamefont {Mark}\ \bibnamefont {Trodden}},\ }\bibfield  {title}
  {\enquote {\bibinfo {title} {{Chameleon early dark energy and the Hubble
  tension}},}\ }\href {\doibase 10.1103/PhysRevD.105.063535} {\bibfield
  {journal} {\bibinfo  {journal} {Phys. Rev. D}\ }\textbf {\bibinfo {volume}
  {105}},\ \bibinfo {pages} {063535} (\bibinfo {year} {2022})},\ \Eprint
  {http://arxiv.org/abs/2106.13290} {arXiv:2106.13290 [astro-ph.CO]}
  \BibitemShut {NoStop}%
\bibitem [{\citenamefont {McDonough}\ \emph {et~al.}(2022)\citenamefont
  {McDonough}, \citenamefont {Lin}, \citenamefont {Hill}, \citenamefont {Hu},\
  and\ \citenamefont {Zhou}}]{McDonough:2021pdg}%
  \BibitemOpen
  \bibfield  {author} {\bibinfo {author} {\bibfnamefont {Evan}\ \bibnamefont
  {McDonough}}, \bibinfo {author} {\bibfnamefont {Meng-Xiang}\ \bibnamefont
  {Lin}}, \bibinfo {author} {\bibfnamefont {J.~Colin}\ \bibnamefont {Hill}},
  \bibinfo {author} {\bibfnamefont {Wayne}\ \bibnamefont {Hu}}, \ and\ \bibinfo
  {author} {\bibfnamefont {Shengjia}\ \bibnamefont {Zhou}},\ }\bibfield
  {title} {\enquote {\bibinfo {title} {{Early dark sector, the Hubble tension,
  and the swampland}},}\ }\href {\doibase 10.1103/PhysRevD.106.043525}
  {\bibfield  {journal} {\bibinfo  {journal} {Phys. Rev. D}\ }\textbf {\bibinfo
  {volume} {106}},\ \bibinfo {pages} {043525} (\bibinfo {year} {2022})},\
  \Eprint {http://arxiv.org/abs/2112.09128} {arXiv:2112.09128 [astro-ph.CO]}
  \BibitemShut {NoStop}%
\bibitem [{\citenamefont {Aghanim}\ \emph
  {et~al.}(2020{\natexlab{c}})\citenamefont {Aghanim} \emph
  {et~al.}}]{Planck:2019nip}%
  \BibitemOpen
  \bibfield  {author} {\bibinfo {author} {\bibfnamefont {N.}~\bibnamefont
  {Aghanim}} \emph {et~al.} (\bibinfo {collaboration} {Planck}),\ }\bibfield
  {title} {\enquote {\bibinfo {title} {{Planck 2018 results. V. CMB power
  spectra and likelihoods}},}\ }\href {\doibase 10.1051/0004-6361/201936386}
  {\bibfield  {journal} {\bibinfo  {journal} {Astron. Astrophys.}\ }\textbf
  {\bibinfo {volume} {641}},\ \bibinfo {pages} {A5} (\bibinfo {year}
  {2020}{\natexlab{c}})},\ \Eprint {http://arxiv.org/abs/1907.12875}
  {arXiv:1907.12875 [astro-ph.CO]} \BibitemShut {NoStop}%
\bibitem [{\citenamefont {Aghanim}\ \emph
  {et~al.}(2020{\natexlab{d}})\citenamefont {Aghanim} \emph
  {et~al.}}]{Planck:2018lbu}%
  \BibitemOpen
  \bibfield  {author} {\bibinfo {author} {\bibfnamefont {N.}~\bibnamefont
  {Aghanim}} \emph {et~al.} (\bibinfo {collaboration} {Planck}),\ }\bibfield
  {title} {\enquote {\bibinfo {title} {{Planck 2018 results. VIII.
  Gravitational lensing}},}\ }\href {\doibase 10.1051/0004-6361/201833886}
  {\bibfield  {journal} {\bibinfo  {journal} {Astron. Astrophys.}\ }\textbf
  {\bibinfo {volume} {641}},\ \bibinfo {pages} {A8} (\bibinfo {year}
  {2020}{\natexlab{d}})},\ \Eprint {http://arxiv.org/abs/1807.06210}
  {arXiv:1807.06210 [astro-ph.CO]} \BibitemShut {NoStop}%
\bibitem [{\citenamefont {Scolnic}\ \emph {et~al.}(2018)\citenamefont {Scolnic}
  \emph {et~al.}}]{Pan-STARRS1:2017jku}%
  \BibitemOpen
  \bibfield  {author} {\bibinfo {author} {\bibfnamefont {D.~M.}\ \bibnamefont
  {Scolnic}} \emph {et~al.} (\bibinfo {collaboration} {Pan-STARRS1}),\
  }\bibfield  {title} {\enquote {\bibinfo {title} {{The Complete Light-curve
  Sample of Spectroscopically Confirmed SNe Ia from Pan-STARRS1 and
  Cosmological Constraints from the Combined Pantheon Sample}},}\ }\href
  {\doibase 10.3847/1538-4357/aab9bb} {\bibfield  {journal} {\bibinfo
  {journal} {Astrophys. J.}\ }\textbf {\bibinfo {volume} {859}},\ \bibinfo
  {pages} {101} (\bibinfo {year} {2018})},\ \Eprint
  {http://arxiv.org/abs/1710.00845} {arXiv:1710.00845 [astro-ph.CO]}
  \BibitemShut {NoStop}%
\bibitem [{\citenamefont {Beutler}\ \emph {et~al.}(2011)\citenamefont
  {Beutler}, \citenamefont {Blake}, \citenamefont {Colless}, \citenamefont
  {Jones}, \citenamefont {Staveley-Smith}, \citenamefont {Campbell},
  \citenamefont {Parker}, \citenamefont {Saunders},\ and\ \citenamefont
  {Watson}}]{Beutler:2011hx}%
  \BibitemOpen
  \bibfield  {author} {\bibinfo {author} {\bibfnamefont {Florian}\ \bibnamefont
  {Beutler}}, \bibinfo {author} {\bibfnamefont {Chris}\ \bibnamefont {Blake}},
  \bibinfo {author} {\bibfnamefont {Matthew}\ \bibnamefont {Colless}}, \bibinfo
  {author} {\bibfnamefont {D.~Heath}\ \bibnamefont {Jones}}, \bibinfo {author}
  {\bibfnamefont {Lister}\ \bibnamefont {Staveley-Smith}}, \bibinfo {author}
  {\bibfnamefont {Lachlan}\ \bibnamefont {Campbell}}, \bibinfo {author}
  {\bibfnamefont {Quentin}\ \bibnamefont {Parker}}, \bibinfo {author}
  {\bibfnamefont {Will}\ \bibnamefont {Saunders}}, \ and\ \bibinfo {author}
  {\bibfnamefont {Fred}\ \bibnamefont {Watson}},\ }\bibfield  {title} {\enquote
  {\bibinfo {title} {{The 6dF Galaxy Survey: Baryon Acoustic Oscillations and
  the Local Hubble Constant}},}\ }\href {\doibase
  10.1111/j.1365-2966.2011.19250.x} {\bibfield  {journal} {\bibinfo  {journal}
  {Mon. Not. Roy. Astron. Soc.}\ }\textbf {\bibinfo {volume} {416}},\ \bibinfo
  {pages} {3017--3032} (\bibinfo {year} {2011})},\ \Eprint
  {http://arxiv.org/abs/1106.3366} {arXiv:1106.3366 [astro-ph.CO]} \BibitemShut
  {NoStop}%
\bibitem [{\citenamefont {Ross}\ \emph {et~al.}(2015)\citenamefont {Ross},
  \citenamefont {Samushia}, \citenamefont {Howlett}, \citenamefont {Percival},
  \citenamefont {Burden},\ and\ \citenamefont {Manera}}]{Ross:2014qpa}%
  \BibitemOpen
  \bibfield  {author} {\bibinfo {author} {\bibfnamefont {Ashley~J.}\
  \bibnamefont {Ross}}, \bibinfo {author} {\bibfnamefont {Lado}\ \bibnamefont
  {Samushia}}, \bibinfo {author} {\bibfnamefont {Cullan}\ \bibnamefont
  {Howlett}}, \bibinfo {author} {\bibfnamefont {Will~J.}\ \bibnamefont
  {Percival}}, \bibinfo {author} {\bibfnamefont {Angela}\ \bibnamefont
  {Burden}}, \ and\ \bibinfo {author} {\bibfnamefont {Marc}\ \bibnamefont
  {Manera}},\ }\bibfield  {title} {\enquote {\bibinfo {title} {{The clustering
  of the SDSS DR7 main Galaxy sample \textendash{} I. A 4 per cent distance
  measure at $z = 0.15$}},}\ }\href {\doibase 10.1093/mnras/stv154} {\bibfield
  {journal} {\bibinfo  {journal} {Mon. Not. Roy. Astron. Soc.}\ }\textbf
  {\bibinfo {volume} {449}},\ \bibinfo {pages} {835--847} (\bibinfo {year}
  {2015})},\ \Eprint {http://arxiv.org/abs/1409.3242} {arXiv:1409.3242
  [astro-ph.CO]} \BibitemShut {NoStop}%
\bibitem [{\citenamefont {Alam}\ \emph {et~al.}(2017)\citenamefont {Alam} \emph
  {et~al.}}]{BOSS:2016wmc}%
  \BibitemOpen
  \bibfield  {author} {\bibinfo {author} {\bibfnamefont {Shadab}\ \bibnamefont
  {Alam}} \emph {et~al.} (\bibinfo {collaboration} {BOSS}),\ }\bibfield
  {title} {\enquote {\bibinfo {title} {{The clustering of galaxies in the
  completed SDSS-III Baryon Oscillation Spectroscopic Survey: cosmological
  analysis of the DR12 galaxy sample}},}\ }\href {\doibase
  10.1093/mnras/stx721} {\bibfield  {journal} {\bibinfo  {journal} {Mon. Not.
  Roy. Astron. Soc.}\ }\textbf {\bibinfo {volume} {470}},\ \bibinfo {pages}
  {2617--2652} (\bibinfo {year} {2017})},\ \Eprint
  {http://arxiv.org/abs/1607.03155} {arXiv:1607.03155 [astro-ph.CO]}
  \BibitemShut {NoStop}%
\bibitem [{\citenamefont {Torrado}\ and\ \citenamefont
  {Lewis}(2021)}]{Torrado:2020dgo}%
  \BibitemOpen
  \bibfield  {author} {\bibinfo {author} {\bibfnamefont {Jesus}\ \bibnamefont
  {Torrado}}\ and\ \bibinfo {author} {\bibfnamefont {Antony}\ \bibnamefont
  {Lewis}},\ }\bibfield  {title} {\enquote {\bibinfo {title} {{Cobaya: Code for
  Bayesian Analysis of hierarchical physical models}},}\ }\href {\doibase
  10.1088/1475-7516/2021/05/057} {\bibfield  {journal} {\bibinfo  {journal}
  {JCAP}\ }\textbf {\bibinfo {volume} {05}},\ \bibinfo {pages} {057} (\bibinfo
  {year} {2021})},\ \Eprint {http://arxiv.org/abs/2005.05290} {arXiv:2005.05290
  [astro-ph.IM]} \BibitemShut {NoStop}%
\bibitem [{\citenamefont {Gelman}\ and\ \citenamefont
  {Rubin}(1992)}]{Gelman:1992zz}%
  \BibitemOpen
  \bibfield  {author} {\bibinfo {author} {\bibfnamefont {Andrew}\ \bibnamefont
  {Gelman}}\ and\ \bibinfo {author} {\bibfnamefont {Donald~B.}\ \bibnamefont
  {Rubin}},\ }\bibfield  {title} {\enquote {\bibinfo {title} {{Inference from
  Iterative Simulation Using Multiple Sequences}},}\ }\href {\doibase
  10.1214/ss/1177011136} {\bibfield  {journal} {\bibinfo  {journal} {Statist.
  Sci.}\ }\textbf {\bibinfo {volume} {7}},\ \bibinfo {pages} {457--472}
  (\bibinfo {year} {1992})}\BibitemShut {NoStop}%
\bibitem [{\citenamefont {Hill}\ \emph {et~al.}(2020)\citenamefont {Hill},
  \citenamefont {McDonough}, \citenamefont {Toomey},\ and\ \citenamefont
  {Alexander}}]{Hill:2020osr}%
  \BibitemOpen
  \bibfield  {author} {\bibinfo {author} {\bibfnamefont {J.~Colin}\
  \bibnamefont {Hill}}, \bibinfo {author} {\bibfnamefont {Evan}\ \bibnamefont
  {McDonough}}, \bibinfo {author} {\bibfnamefont {Michael~W.}\ \bibnamefont
  {Toomey}}, \ and\ \bibinfo {author} {\bibfnamefont {Stephon}\ \bibnamefont
  {Alexander}},\ }\bibfield  {title} {\enquote {\bibinfo {title} {{Early dark
  energy does not restore cosmological concordance}},}\ }\href {\doibase
  10.1103/PhysRevD.102.043507} {\bibfield  {journal} {\bibinfo  {journal}
  {Phys. Rev. D}\ }\textbf {\bibinfo {volume} {102}},\ \bibinfo {pages}
  {043507} (\bibinfo {year} {2020})},\ \Eprint
  {http://arxiv.org/abs/2003.07355} {arXiv:2003.07355 [astro-ph.CO]}
  \BibitemShut {NoStop}%
\bibitem [{\citenamefont {Murgia}\ \emph {et~al.}(2021)\citenamefont {Murgia},
  \citenamefont {Abell\'an},\ and\ \citenamefont {Poulin}}]{Murgia:2020ryi}%
  \BibitemOpen
  \bibfield  {author} {\bibinfo {author} {\bibfnamefont {Riccardo}\
  \bibnamefont {Murgia}}, \bibinfo {author} {\bibfnamefont {Guillermo~F.}\
  \bibnamefont {Abell\'an}}, \ and\ \bibinfo {author} {\bibfnamefont {Vivian}\
  \bibnamefont {Poulin}},\ }\bibfield  {title} {\enquote {\bibinfo {title}
  {{Early dark energy resolution to the Hubble tension in light of weak lensing
  surveys and lensing anomalies}},}\ }\href {\doibase
  10.1103/PhysRevD.103.063502} {\bibfield  {journal} {\bibinfo  {journal}
  {Phys. Rev. D}\ }\textbf {\bibinfo {volume} {103}},\ \bibinfo {pages}
  {063502} (\bibinfo {year} {2021})},\ \Eprint
  {http://arxiv.org/abs/2009.10733} {arXiv:2009.10733 [astro-ph.CO]}
  \BibitemShut {NoStop}%
\bibitem [{\citenamefont {Ivanov}\ \emph {et~al.}(2020)\citenamefont {Ivanov},
  \citenamefont {McDonough}, \citenamefont {Hill}, \citenamefont {Simonovi\'c},
  \citenamefont {Toomey}, \citenamefont {Alexander},\ and\ \citenamefont
  {Zaldarriaga}}]{Ivanov:2020ril}%
  \BibitemOpen
  \bibfield  {author} {\bibinfo {author} {\bibfnamefont {Mikhail~M.}\
  \bibnamefont {Ivanov}}, \bibinfo {author} {\bibfnamefont {Evan}\ \bibnamefont
  {McDonough}}, \bibinfo {author} {\bibfnamefont {J.~Colin}\ \bibnamefont
  {Hill}}, \bibinfo {author} {\bibfnamefont {Marko}\ \bibnamefont
  {Simonovi\'c}}, \bibinfo {author} {\bibfnamefont {Michael~W.}\ \bibnamefont
  {Toomey}}, \bibinfo {author} {\bibfnamefont {Stephon}\ \bibnamefont
  {Alexander}}, \ and\ \bibinfo {author} {\bibfnamefont {Matias}\ \bibnamefont
  {Zaldarriaga}},\ }\bibfield  {title} {\enquote {\bibinfo {title}
  {{Constraining Early Dark Energy with Large-Scale Structure}},}\ }\href
  {\doibase 10.1103/PhysRevD.102.103502} {\bibfield  {journal} {\bibinfo
  {journal} {Phys. Rev. D}\ }\textbf {\bibinfo {volume} {102}},\ \bibinfo
  {pages} {103502} (\bibinfo {year} {2020})},\ \Eprint
  {http://arxiv.org/abs/2006.11235} {arXiv:2006.11235 [astro-ph.CO]}
  \BibitemShut {NoStop}%
\bibitem [{\citenamefont {Jiang}\ and\ \citenamefont
  {Piao}(2021)}]{Jiang:2021bab}%
  \BibitemOpen
  \bibfield  {author} {\bibinfo {author} {\bibfnamefont {Jun-Qian}\
  \bibnamefont {Jiang}}\ and\ \bibinfo {author} {\bibfnamefont {Yun-Song}\
  \bibnamefont {Piao}},\ }\bibfield  {title} {\enquote {\bibinfo {title}
  {{Testing AdS early dark energy with Planck, SPTpol, and LSS data}},}\ }\href
  {\doibase 10.1103/PhysRevD.104.103524} {\bibfield  {journal} {\bibinfo
  {journal} {Phys. Rev. D}\ }\textbf {\bibinfo {volume} {104}},\ \bibinfo
  {pages} {103524} (\bibinfo {year} {2021})},\ \Eprint
  {http://arxiv.org/abs/2107.07128} {arXiv:2107.07128 [astro-ph.CO]}
  \BibitemShut {NoStop}%
\bibitem [{\citenamefont {Ye}\ \emph {et~al.}(2021)\citenamefont {Ye},
  \citenamefont {Hu},\ and\ \citenamefont {Piao}}]{Ye:2021nej}%
  \BibitemOpen
  \bibfield  {author} {\bibinfo {author} {\bibfnamefont {Gen}\ \bibnamefont
  {Ye}}, \bibinfo {author} {\bibfnamefont {Bin}\ \bibnamefont {Hu}}, \ and\
  \bibinfo {author} {\bibfnamefont {Yun-Song}\ \bibnamefont {Piao}},\
  }\bibfield  {title} {\enquote {\bibinfo {title} {{Implication of the Hubble
  tension for the primordial Universe in light of recent cosmological data}},}\
  }\href {\doibase 10.1103/PhysRevD.104.063510} {\bibfield  {journal} {\bibinfo
   {journal} {Phys. Rev. D}\ }\textbf {\bibinfo {volume} {104}},\ \bibinfo
  {pages} {063510} (\bibinfo {year} {2021})},\ \Eprint
  {http://arxiv.org/abs/2103.09729} {arXiv:2103.09729 [astro-ph.CO]}
  \BibitemShut {NoStop}%
\bibitem [{\citenamefont {Ye}\ and\ \citenamefont {Piao}(2022)}]{Ye:2022afu}%
  \BibitemOpen
  \bibfield  {author} {\bibinfo {author} {\bibfnamefont {Gen}\ \bibnamefont
  {Ye}}\ and\ \bibinfo {author} {\bibfnamefont {Yun-Song}\ \bibnamefont
  {Piao}},\ }\bibfield  {title} {\enquote {\bibinfo {title} {{Improved
  constraints on primordial gravitational waves in light of the H0 tension and
  BICEP/Keck data}},}\ }\href {\doibase 10.1103/PhysRevD.106.043536} {\bibfield
   {journal} {\bibinfo  {journal} {Phys. Rev. D}\ }\textbf {\bibinfo {volume}
  {106}},\ \bibinfo {pages} {043536} (\bibinfo {year} {2022})},\ \Eprint
  {http://arxiv.org/abs/2202.10055} {arXiv:2202.10055 [astro-ph.CO]}
  \BibitemShut {NoStop}%
\bibitem [{\citenamefont {Xu}\ and\ \citenamefont {Huang}(2018)}]{Xu:2016ddc}%
  \BibitemOpen
  \bibfield  {author} {\bibinfo {author} {\bibfnamefont {Lixin}\ \bibnamefont
  {Xu}}\ and\ \bibinfo {author} {\bibfnamefont {Qing-Guo}\ \bibnamefont
  {Huang}},\ }\bibfield  {title} {\enquote {\bibinfo {title} {{Detecting the
  Neutrinos Mass Hierarchy from Cosmological Data}},}\ }\href {\doibase
  10.1007/s11433-017-9125-0} {\bibfield  {journal} {\bibinfo  {journal} {Sci.
  China Phys. Mech. Astron.}\ }\textbf {\bibinfo {volume} {61}},\ \bibinfo
  {pages} {039521} (\bibinfo {year} {2018})},\ \Eprint
  {http://arxiv.org/abs/1611.05178} {arXiv:1611.05178 [astro-ph.CO]}
  \BibitemShut {NoStop}%
\bibitem [{\citenamefont {Lewis}(2019)}]{Lewis:2019xzd}%
  \BibitemOpen
  \bibfield  {author} {\bibinfo {author} {\bibfnamefont {Antony}\ \bibnamefont
  {Lewis}},\ }\bibfield  {title} {\enquote {\bibinfo {title} {{GetDist: a
  Python package for analysing Monte Carlo samples}},}\ }\href@noop {} {\
  (\bibinfo {year} {2019})},\ \Eprint {http://arxiv.org/abs/1910.13970}
  {arXiv:1910.13970 [astro-ph.IM]} \BibitemShut {NoStop}%
\end{thebibliography}%
\end{document}